\documentclass[iop, numberedappendix, appendixfloats, manuscript, tighten]{aastex}
\usepackage[]{times, graphicx}
\usepackage{float}
 \newcommand{\pref}{\protect\ref}

 \newcommand{\soho}{{\em SOHO}}
 \newcommand{\sdo}{{\em SDO}}
 \newcommand{\ace}{{\em ACE}}
 \newcommand{\ulysses}{{\em Ulysses}}

\newcommand{\qfe}{$\langle \mathrm{Q}_{{\mathrm{Fe}}}\rangle$}
\newcommand{\vsw}{$\mathrm{V}_{SW}$}
\newcommand{\ahe}{$\mathrm{A}_{{\mathrm{He}}}$}
 
\begin{document}
 
\shorttitle{Hemispheric Asymmetries in Activity}
\shortauthors{S.~W. McIntosh et~al.}
\title{Hemispheric Asymmetries of Solar Photospheric Magnetism: Radiative, Particulate, and Heliospheric Impacts}
\author{Scott W. McIntosh\altaffilmark{1}, Robert J. Leamon\altaffilmark{2}, Joseph B. Gurman\altaffilmark{3}, Jean-Philippe Olive\altaffilmark{4}, Jonathan W. Cirtain\altaffilmark{5}, David H. Hathaway\altaffilmark{5}, Joan Burkepile\altaffilmark{1}, Mark Miesch\altaffilmark{1}, Robert S. Markel\altaffilmark{1}, Leonard Sitongia\altaffilmark{1}}
\altaffiltext{1}{High Altitude Observatory, National Center for Atmospheric Research, P.O. Box 3000, Boulder CO 80307}
\altaffiltext{2}{Department of Physics, Montana State University, Bozeman, MT 59717}
\altaffiltext{3}{Solar Physics Laboratory, NASA Goddard Space Flight Center, Greenbelt MD 20771}
\altaffiltext{4}{Astrium SAS, 6 rue Laurent Pichat, 75016 Paris, France}
\altaffiltext{5}{Marshall Space Flight Center, Huntsville, AL 35812}

\begin{abstract}
Among many other measurable quantities the summer of 2009 saw a considerable low in the radiative output of the Sun that was temporally coincident with the largest cosmic ray flux ever measured at 1AU. Combining measurements and observations made by the Solar and Heliospheric Observatory (SOHO) and Solar Dynamics Observatory (SDO) spacecraft we begin to explore the complexities of the descending phase of solar cycle 23, through the 2009 minimum into the ascending phase of solar cycle 24. A hemispheric asymmetry in magnetic activity is clearly observed and its evolution monitored and the resulting (prolonged) magnetic imbalance must have had a considerable impact on the structure and energetics of the heliosphere. While we cannot uniquely tie the variance and scale of the surface magnetism to the dwindling radiative and particulate output of the star, or the increased cosmic ray flux through the 2009 minimum, the timing of the decline and rapid recovery in early 2010 would appear to inextricably link them. These observations support a picture where the SunÕs hemispheres are significantly out of phase with each other. Studying historical sunspot records with this picture in mind shows that the northern hemisphere has been leading since the middle of the last century and that the hemispheric ÒdominanceÓ has changed twice in the past 130 years. The observations presented give clear cause for concern, especially with respect to our present understanding of the processes that produce the surface magnetism in the (hidden) solar interior - hemispheric asymmetry is the normal state - the strong symmetry shown in 1996 was abnormal. Further, these observations show that the mechanism(s) which create and transport the magnetic flux magnetic flux are slowly changing with time and, it appears, with only loose coupling across the equator such that those asymmetries can persist for a considerable time. As the current asymmetry persists and the basal energetics of the system continue to dwindle we anticipate new radiative and particulate lows coupled with increased cosmic ray fluxes heading into the next solar minimum.
\end{abstract}
 
 
\section{Introduction}
The most recent solar minimum has drawn a great deal of interest \citep[e.g., ][]{2008GeoRL..3518103M, 2009JGRA..11409105G, 2009SoPh..257...99K, 2010ApJ...723L...1M, 2010arXiv1009.0784P, 2010ASPC..428..275G, 2010GeoRL..3716103S, 2011SoPh..tmp...24A, 2011ApJ...736..136W, 2011ApJ...730L...3M, 2011ApJ...740L..23M}, but largely because it has not behaved in this fashion during the epoch of routine detailed observation of surface and sub-surface phenomena. Further, there was a belief that we (somehow) understood the basal modulation of the SunÕs radiative and particulate output and that it was invariant; that is not the caseÑno two solar minima are the same, we have known this, but the Sun ably demonstrated it for us over the past few years. In our pursuit of a physical understanding of the SunÕs impact on Earth (and the Solar System) we have (correctly) been drawn to processes that relentlessly form magnetic flux in the interior \citep[e.g.,][]{1999ApJ...518..508D, 2005LRSP....2....2C, 2007AdSpR..40..899C, 2007AdSpR..40.1917C, 2009SSRv..144...87B, 2010LRSP....7....1H, 2011Natur.471...80N} and those that govern the eventual eruption of that magnetic flux through the photosphere into the portions of the atmosphere which we can directly observe. The tail of those processes in the outer solar atmosphere, from the perspective of energy density, dictate the output of our star, but our historical reference point has been limited to observing gross (readily observable) proxies of magnetic evolution. It is likely that these proxies carry differently weighed information about our starÕs evolution and it is our task to know the meaningful information content of each, beyond zeroth order application, if we are to move towards comprehensive modeling of the solar atmosphere as a system.

In this Paper we consider several records of quiescent activity over the past solar cycle using measurements from the \soho{}, \sdo{}, \ace{}, and \ulysses{} spacecraft. We pay particular attention to the line-of-sight (LOS) magnetogram data from the \soho{} Michelson Doppler Imager \citep[MDI;][]{1995SoPh..162..129S} to investigate trends in the evolution of surface magnetism that may point to complexities in longer term processes and how we consider/model them. 

In \S\pref{SMV} we demonstrate the minimum-minimum modulation of several Solar outputs and their connection to the relentlessly evolving surface magnetism that motivated this study. Indeed, we infer that these subtle changes may have a broader impact --- on the structure of the heliosphere
itself (\S\pref{CRF}). Studying the MDI record of magnetograms in \S\pref{MDI} we look for gross trends of surface magnetism that may underly these varying quantities. The behavior that we identify in the variation of the magnetogram record is subtle, but clear, and spawns the investigation into
other measures of outer atmospheric structure and small-scale activity in \S\pref{sscale} in search consistent behavior in the hypothesis drawnÑthat the magnetic activity of the northern and southern hemispheres are not in phase with each other. Extending our analysis to a longer sunspot record
(differentiated by hemisphere; \S\pref{sunspot}) we see that this behavior is not isolated to the recent solar minimum but has been visible for a considerable time. We discuss, in the closing sections, that the observed variance in surface magnetism must place stronger physical constraints on the modeling of interior magnetic flux generation and how that evolving flux governs the energetic output of the star as well as the mould of the heliosphere itself. There are strong indications that the two hemispheres are nearly two years out of phase at the time of writing, and it is unclear what the next descending phase and solar minimum will offer, far less those going further into the future --- what evolutionary path the Sun will take next? It is likely that only a significant increase in complexity of modeling can help us establish the real knock-on effects of what can happen to reverse the hemispheric asymmetry and if that is connected to the slowly decreasing levels of solar activity.

\section{2009 Solar Minimum Measurements}\label{SMV}
\begin{figure}
\epsscale{0.5}
\plotone{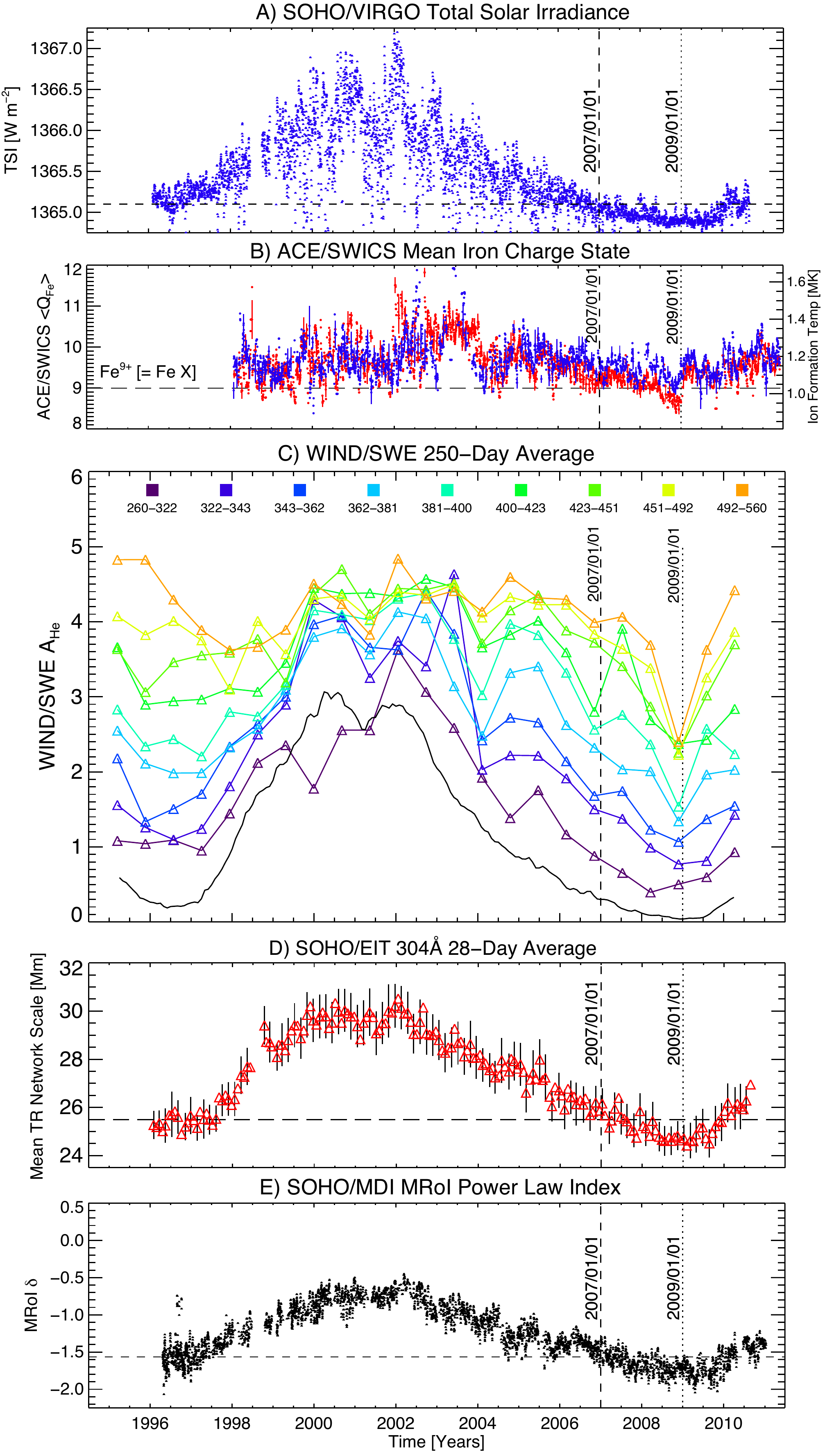}
\caption{From top to bottom the total solar irradiance measured by \soho/VIRGO, the mean Fe charge state (\qfe) measured by \ace/SWICS, the abundance of Helium (\ahe) measured by \ace{}, the apparent scale length of the \soho/EIT transition region network emission, and the power-law index of the \soho/MDI ``Magnetic Range of Influence'' (MRoI over solar cycle 23 and into the early part of cycle 24. For reference we show 2 dates: January~1, 2007 (dashed vertical line) and January~1, 2009 (dotted vertical line) and horizontal dashed lines to indicate the 1996 level of each measurement. The solar wind measures are differentiated by speed as indicated---note that there is no measurement of \qfe{} before 1998. \label{f1}} 
\end{figure}

Figure~\pref{f1} shows the variation of several measures of different origin over the past solar cycle. From top to bottom the figure shows the total solar irradiance (TSI) measured by \soho/VIRGO, the mean Fe charge state (\qfe) measured by \ace/SWICS \citep[][]{1992A&AS...92..267G, 1998SSRv...86....1S}, the abundance of Helium (\ahe) measured by {\em WIND} \citep[e.g.,][]{2001GeoRL..28.2767A,2007ApJ...660..901K}, the apparent scale length of the transition region network emission observed by \soho/EIT, and the power-law index of the MRoI inferred (see below) from \soho/MDI magnetograms over solar cycle 23 and into the early part of cycle 24 \citep[][]{2011ApJ...730L...3M}. The bottom pair of panels illustrate the variation in the separation distance of magnetic field elements in the photosphere and the prevalent ``network'' emission length scale in the quiescent solar atmosphere \citep[][]{2011ApJ...730L...3M}---their variation is remarkably consistent and demonstrates the ``imprint'' of the underlying magnetism on the radiative and particulate output of our star \citep[][]{2011ApJ...739....9S, 2011ApJ...740L..23M, 2012ApJ...745..162K}. For reference we show 2 dates: January~1, 2007 (dashed vertical line) and January~1, 2009 (dotted vertical line) and horizontal dashed lines to indicate the 1996 level of each measurement. 

\subsection{The 2009 High in Cosmic Ray Flux Measured At Earth: Coincidence or Connection?}\label{CRF}
At almost exactly the same time that the energetic and magnetic proxies shown in Fig.~\pref{f1} reached new lows the cosmic ray flux (CRF) as measured at Earth (and in interplanetary space) reached an all time high \citep[in the context of the space age][]{2010ApJ...723L...1M}. We wonder if the apparent temporal correlation of the increased CRF and decrease in proxies of solar magnetism are causally related; or if there is an appreciable coupling between the small-scale and global-scale magnetism that affect the energetics {\em and} structure of the heliosphere? 

The CRF is measured on Earth by a network of neutron monitors that detect the decay products arising from the collision of incoming cosmic rays with nitrogen and oxygen molecules in the upper atmosphere. In this study, we have used neutron monitor data (red points) from the Sodankyla Geophysical Observatory of the University of Oulu, Finland (\url{http://cosmicrays.oulu.fi/}). We use data corrected for variations in local atmospheric pressure since count rates of neutrons at the detector are proportional to the number of collision targets in the upper atmosphere. Further, to demonstrate that the anomalous CRF variation observed during this most recent solar minimum is not an artifact produced by the Earth's atmosphere or magnetosphere, we must consider a non-Earthbound monitor of the space particle environment, such as that studied by \citet{2010ApJ...723L...1M}. Such a data set is provided (serendipitously) by the ``Single Event Flag'', or bad-bit count rate of the solid state recorder (SSR) on the {\em Solar and Heliospheric Observatory} \citep[\soho{};][]{1995SoPh..162....1D, 2002ITNS...49.1345H}. The \soho{} SSR data is discussed in more detail in Appendix~\ref{SSU}.

\begin{figure}[!ht]
\epsscale{1.}
\plotone{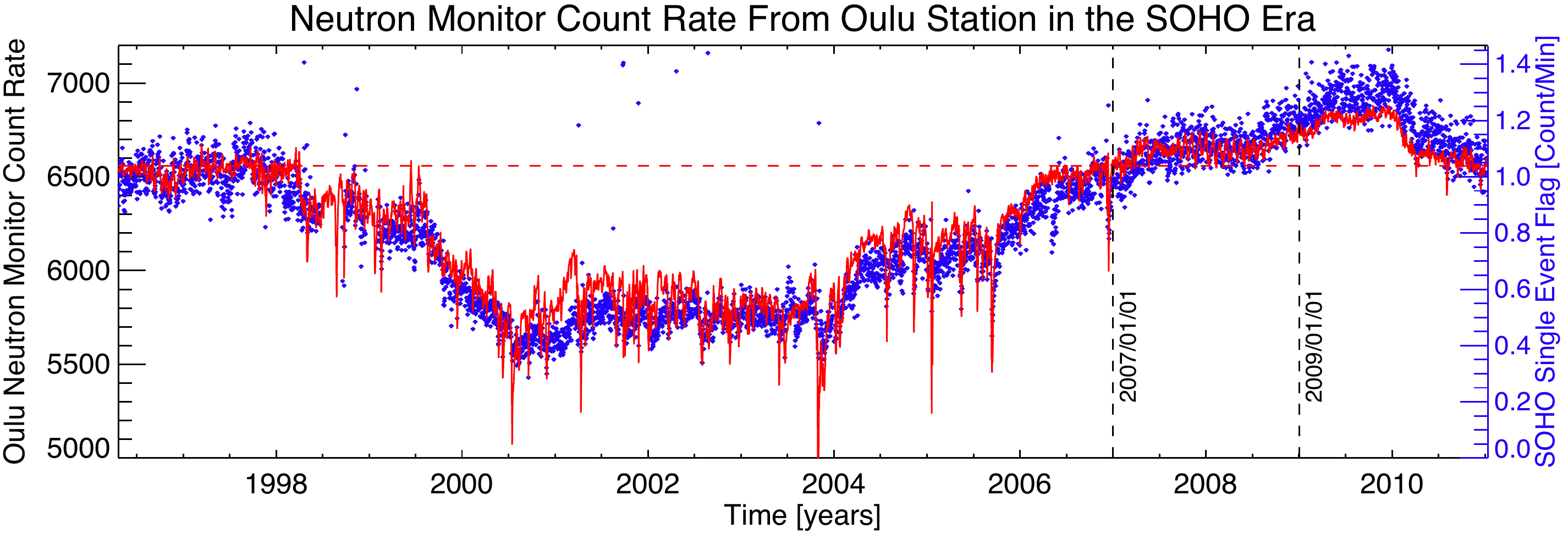}
\caption{Comparing the daily average of the Oulu Station Cosmic Ray Flux (CRF; red) with the SOHO Single Event Flag Count (blue dots) over the period from March 1996 to January 2011 which covers solar cycle~23 (cf. Fig~\pref{f1}). The \soho{} SSR measurements track the variations in the CRF measured at Earth with almost no delay. Like that shown in Fig.~\pref{f1} the dashed vertical black line marks June 1 2008 and indicates the start of the anomalous rise in the CRF. Notice also the apparent low-amplitude 28-day periodicity in the CRF at solar minimum \citep[e.g.,][]{1999ICRC....7..163K,2000SSRv...94...25B}, a detail explored Appendix~\pref{oscill}. \label{f2}} 
\end{figure}

The spacecraft's management software keeps track of the number of bad data bits that are corrected on the SSR, the rate of these corrections is shown in Fig.~\pref{f2} (blue dots). The average rate fluctuates around one bad-bit per minute, ranging from 0.3 to 1.4, plotted on the same graph is the 12-hour average of the Oulu station CRF (red). There are numerous sharp peaks in the timeseries that are driven by solar energetic particle storms, but the general modulation of the bad-bit count rate is driven by changes in basal solar activity cycle that are apparent in the decline in the bad-bit rate as solar maximum is approached. This spacecraft history is correlated well with ground-level neutron monitor data as we observe a remarkable correspondence in the (scaled) variation in range of both measurements over a broad range of timescales, from short flare-related particles, to the longer solar-cycle related variation. In particular, the bad-bit count rate shows a similarly anomalous increase to the CRF, starting in the middle of 2008, and recovering synchronously in late 2009. The approximately 1.5~million km distance of \soho{} from the Earth reinforces the \ace{} measurements \citep{2010ApJ...723L...1M} that the source of the anomaly observed in the Neutron Monitor Stations does not originate at, or near the Earth, but instead results from curious behavior of the Sun over this period.

We draw the reader's attention to the apparent 28-day periodicity visible in the CRF during the solar minima, around late~1996--mid~1997 and again in late~2007 through early~2008. This very low-amplitude, highly modulated variation in the CRF at solar minima is not an isolated phenomenon \citep[e.g.,][]{1999ICRC....7..163K,2000SSRv...94...25B}. The apparent cause of this low-amplitude modulation---a repeating pattern of different of solar magnetic activity on one side of the Sun compared to the other. However, understanding this modulatory phenomenon bears some relevance to the conclusion we will reach below. The response of the CRF to solar surface magnetism fluctuations can be as rapid as only a couple of days, or the time that it takes the fast solar wind traveling at 700~km/s to reach the Earth from the Sun, or the co-rotating interaction region delineating fast and slow wind streams. These are not disturbances in the heliosphere's outer structure, they are directly forced by conditions on the Sun days, and not months, earlier. However, since this topic is not directly relevant to the bulk of the manuscript we defer further discussion to an Appendix (\pref{oscill}).

\section{Cycle 23/24 Magnetogram Trends}\label{MDI}
\soho{} uses the Michelson Doppler Imager \citep[MDI;][]{1995SoPh..162..129S} to probe the interior of the Sun by measuring the photospheric manifestation of solar oscillations and magnetism. The instrument images the Sun on a $1024^{2}$ CCD camera through a series of five narrow spectral filters (FWHM bandwidth of 94~m\AA{}) that span the \ion{Ni}{1} 6768\AA{} spectral line. When polarizers are included in the light path MDI records the line-of-sight (LOS) magnetic field with a resolution of 4\arcsec{} over the whole solar disk using a simple linear combination of the signal in the five sampled line positions \citep[as discussed in][]{1995SoPh..162..129S}.

\begin{figure}[!ht]
\plotone{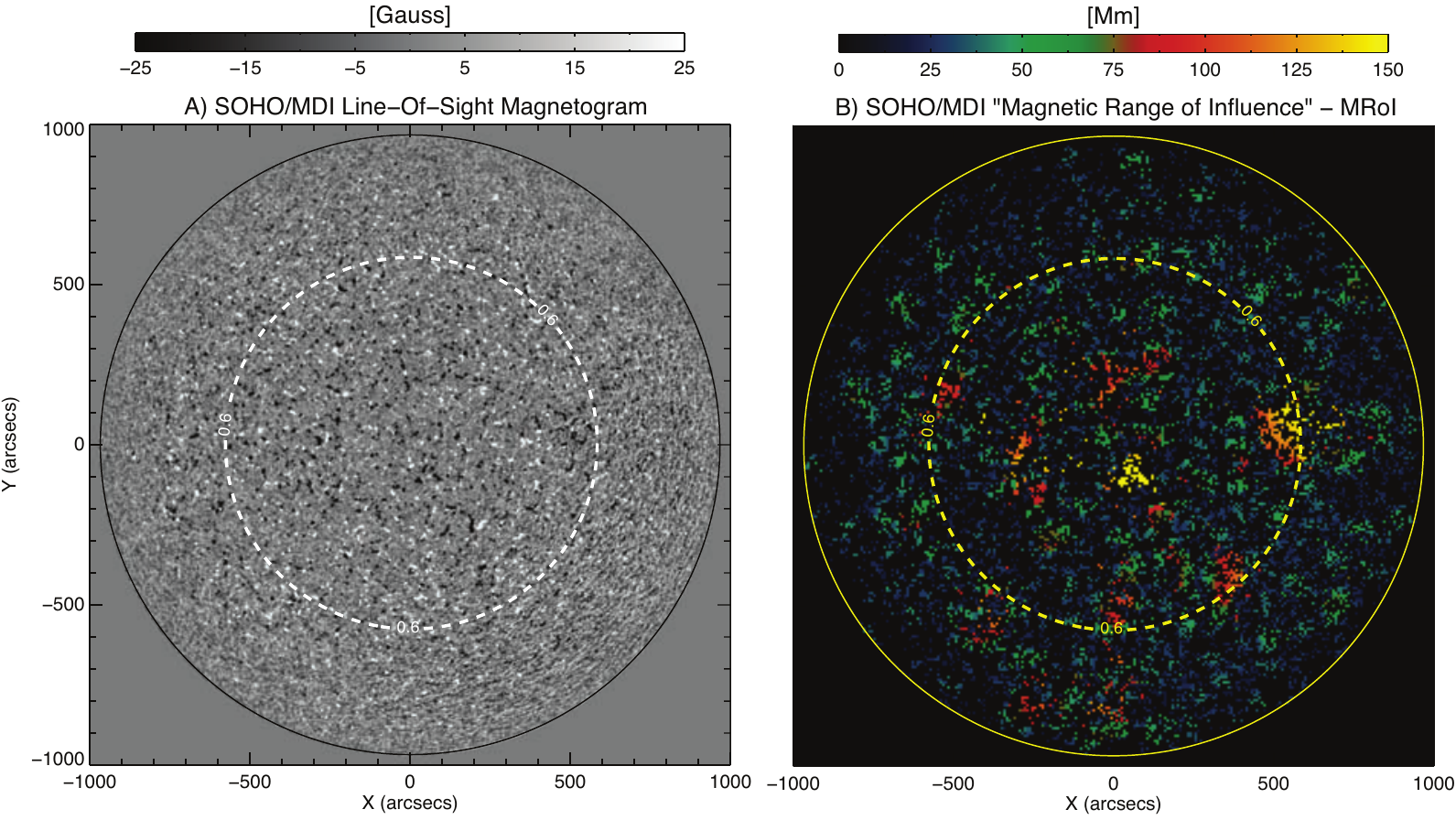}
\caption{The \soho/MDI line-of-sight magnetogram (A) from 00:00UT on April 10 2008 and ``Magnetic Range of Influence'' (B; MRoI) map during the extremely quiet phase near the cycle 23/24 solar minimum. Concentrations of negative magnetic field are black in color while the negative magnetic field are white. The dashed line encircling disk center at a value of $0.6R_\sun$ and outlines the region for which we have performed the magnetogram analysis shown in Fig.~\pref{f4}.\label{f3}} 
\end{figure}

Every 96 minutes MDI takes a synoptic map of the photospheric magnetic field. The example shown (Fig.~\pref{f3}) was taken at 00:00UT on April 10, 2008 when the Sun was very quiet (see, e.g., Fig.~\pref{f3}A). The photosphere, in this case, appears as a mixed ``salt and pepper'' distribution of positive (white) and negative (black) magnetic field concentrations. The first impression of the signal from the disk center area is of a dominance of negative polarity magnetic field. 

The right panel of the figure (Fig.~\pref{f3}B) shows the ``Magnetic Range of Influence'' \citep[MRoI;][]{2006ApJ...644L..87M} for the same magnetogram. The MRoI is the (radial) distance over which we must add magnetic field strengths such that the total of the sum is zero with respect to the central pixel, i.e., it is a crude measure of the connecting length scales in the magnetogram and when MRoI is small the field closes locally and conversely, when it is large, the field would appear to closes a long distance away. It is important to note that all coronal holes and some active regions have large MRoI values.

The dashed line around the center of the solar disk outlines a distance of 60\% of a solar radius and encapsulates the region for which we compute the total signed field (TSF) and total unsigned field (USF) which are defined simply as the total magnetic signal (summing all of the pixels) and the total absolute magnetic signal (summing the absolute value) in that disk center region, respectively. Our rationale for considering only the disk center region is because, at higher latitudes and longitudes, there is a significant uncertainty in accurately determining the LOS magnetic field. In the central region of the solar disk we have the highest degree of confidence in the accuracy of the measurement.

For the analysis shown below we have performed this calculation for every MDI LOS magnetogram of the 96-minute series that has an exposure time of 300s and have the highest signal to noise ratio. There were nearly 30000 of these magnetograms taken by MDI between 1996 April~21 and 2011 February~1 and the noise background is expected to be at the level of 5~Gauss per pixel.

\subsection{The Variation of Spatially Integrated Photospheric Magnetism}
\begin{figure}[!ht]
\plotone{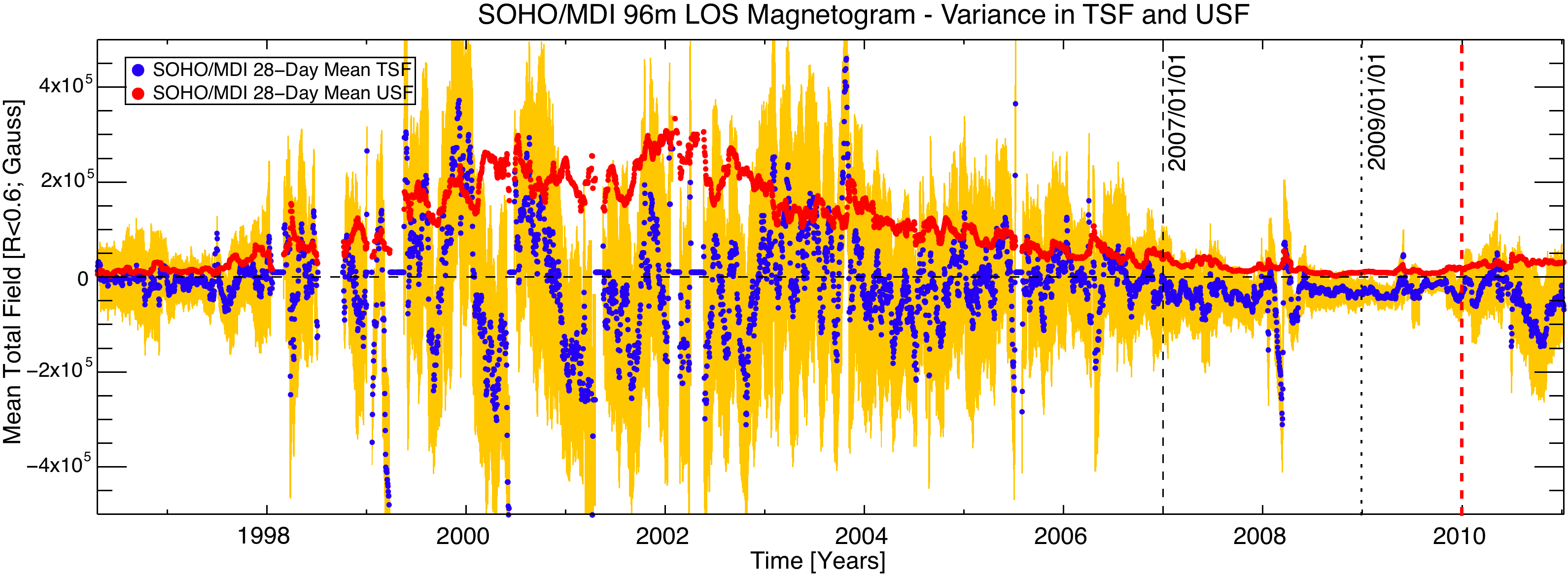}
\caption{The variation of the 28-day running average \soho/MDI measured disk center total signed (TSF; blue dots) and total unsigned (USF; red dots) magnetic field over solar cycle 23 and into cycle 24. The orange band in the plot demonstrates the variance of the 28-day running average TSF.\label{f4}} 
\end{figure}

We consider the variation of the total magnetic field measured in a circular region of $0.6R_\sun$ measured from solar disk center from early 1996 to the start of 2011. In using the LOS magnetograms provided every 96 minutes we minimize systematic variance in the data by routinely mapping the central portion of the solar disk a number of times per day, day after day, for many years. In Figure~\pref{f3} we show the variation of the 28-day running average MDI measured disk center TSF (blue dots) and USF (red dots) magnetic field strength over solar cycle 23 and into cycle 24. The orange band in the plot demonstrates the variance of the 28-day running average TSF. We see that during the time encapsulating the solar minimum of 2009 (2007 January~1--2010 January~1) there is an interval when the TSF is unbalanced and predominantly negative. Further, the TSF is negative by a value of about the variance in the measurement from 2008 June~1 until 2010 January~1, except for a couple of very short excursions. This is in stark contrast to solar minimum conditions in 1996/1997 where the TSF fluctuates about zero with a considerably larger variance. {\em During the 2009 solar minimum the equatorial region of the Sun was imbalanced, with a net negative field. That imbalance persisted for almost two years.}

We should stress that our use of the term ``imbalance'', or later use of the term ``unipolar,''  with reference to the distribution of the photospheric magnetic field implies that the spatial magnetic field distribution is dominated by one polarity over a given scale length---polar coronal holes are an example of a prolonged (local) imbalance and unipolarity of the magnetic field. The use of these terms in no way indicates that the entire solar atmosphere is unipolar or imbalanced (implying a breakdown in the fundamental laws of physics) where local imbalances of one polarity are balanced by local concentrations of the other polarity that are not necessarily nearby.

\subsection{Long-lived Unipolar Regions?}
\begin{figure}[!ht]
\plotone{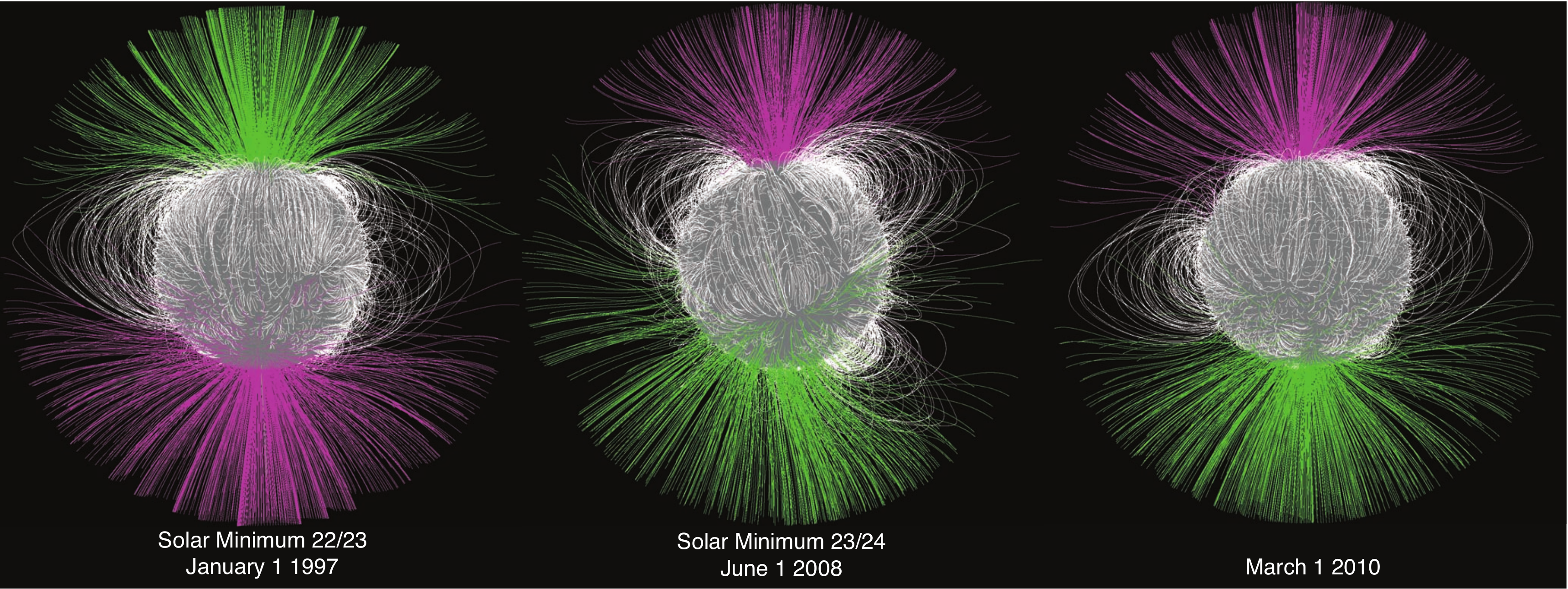}
\caption{Comparing PFSS models of the cycle 22/23 minimum (left) to those around the 2009 solar minimum (center and right). The ``open'' magnetic field lines rooted in regions of negative magnetic polarity are pink while those from positive polarity regions are green. Magnetic field lines that find opposite polarity magnetic fields (i.e., those that are closed locally) are white. \label{f5}} 
\end{figure}

The implication of Fig.~\pref{f4} is that the equatorial region of the solar atmosphere is dominated by one single polarity of magnetic field through the 2009 solar minimum. If this is indeed the case then we should expect some impact of this on the global morphology of Sun's magnetic field. Consider Fig.~\pref{f5}, which shows three visualizations of the coronal magnetic field topology near the 1997 and 2009 solar minima and in the first quarter of 2010 as approximated by the Potential Field Source-Surface (PFSS) model extrapolation of the LOS SOHO/MDI photospheric magnetic field \citep{1969SoPh....6..442S}. Notice that the structure of the 1997 corona (left) is (approximately) symmetric about the solar equator while the 2008 image (center; noting the polarity reversal of the northern and southern poles) shows open magnetic field of negative polarity spans nearly two-thirds of the off-limb corona. As solar cycle 24 begins in earnest (right) we see a coronal environment that has a high degree of north-south symmetry. 

\begin{figure}[!ht]
\plotone{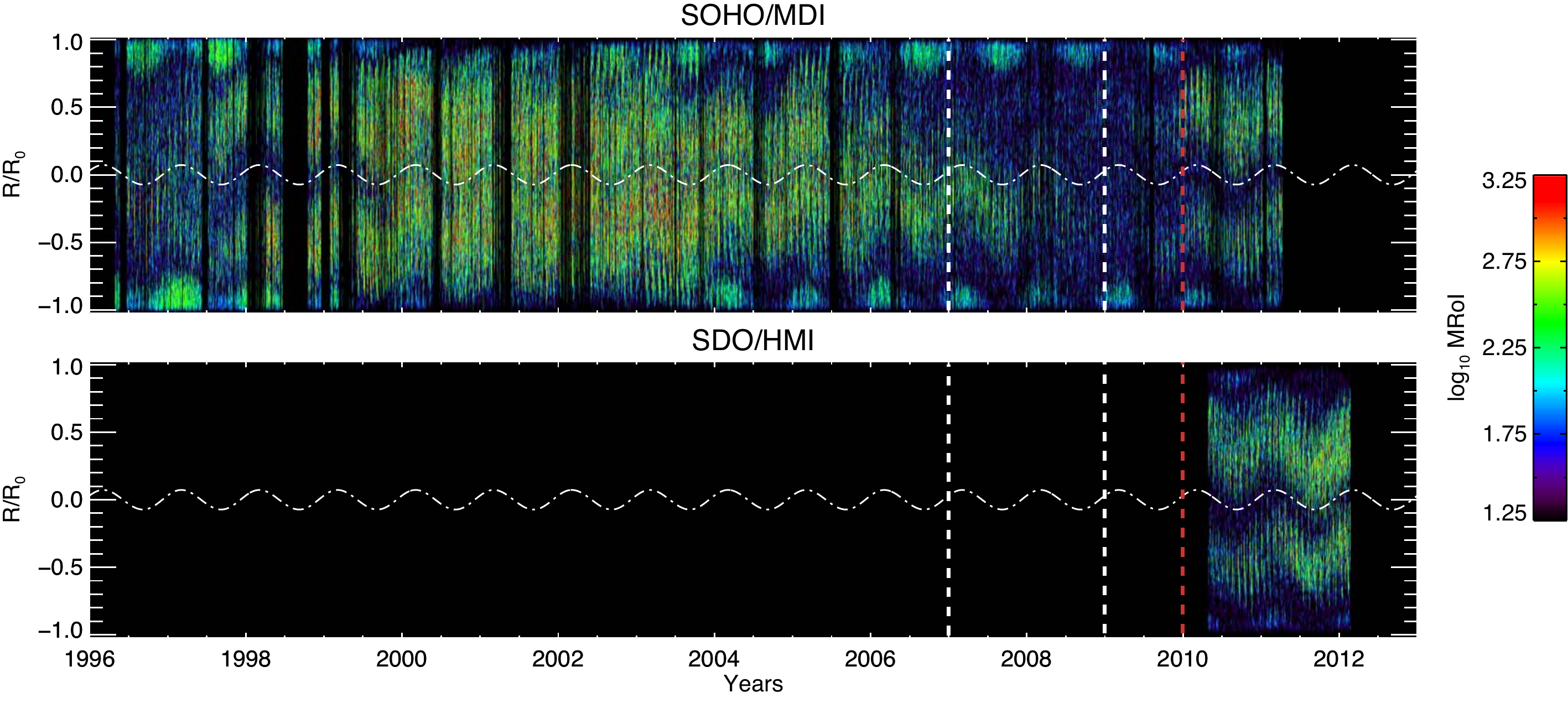}
\caption{The temporal evolution of the ``Magnetic Range of Influence'' (MRoI) around the central meridian using \soho/MDI (top) and \sdo/HMI (bottom) magnetograms. The white sinusoidal dash-dotted line indicates the variation of the Sun's rotational axis (``B\--angle'') over time and gives the reader a feel for the latitude of the solar equator. \label{f6}} 
\end{figure}

Another representation of this over the period of concern is shown in Fig.~\pref{f6} where we show the latitudinal variation of the MRoI (see Fig.~\pref{f1}) for the same magnetograms analyzed in Fig.~\pref{f4} averaged over a 10\arcsec{} wide window about the central meridian. Recalling from above (Sect.~\pref{MDI}) that the MRoI is an indicator of the length over which the magnetic field is locally unbalanced the progression shown in the figure panels (for \soho/MDI and \sdo/HMI) shows several things: the norther hemispheric activity belt seems to disappear in 2006; and the southern hemisphere exhibits large MRoI values well into 2008 which appear to cross the equator\footnote{The rapid change in the northern activity belt is clearly visible in ``butterfly'' diagrams of photospheric magnetograms \citep[e.g., Fig.~1 of][and \url{http://solarscience.msfc.nasa.gov/dynamo.shtml} or \url{http://solarscience.msfc.nasa.gov/greenwch.shtml} for an ongoing record]{Gonzales2011}}. We also see that the northern hemisphere has magnetic features with large MRoI late in 2009 (or early in 2010) as indicated by the red dashed vertical line. This figure would appear to indicate that there is a pronounced difference in the timescales evolving the northern and southern hemispheres, with the latter lagging by an observable amount. 

\subsection{The ``Rush to the Poles''}
\begin{figure}[!ht]
\plotone{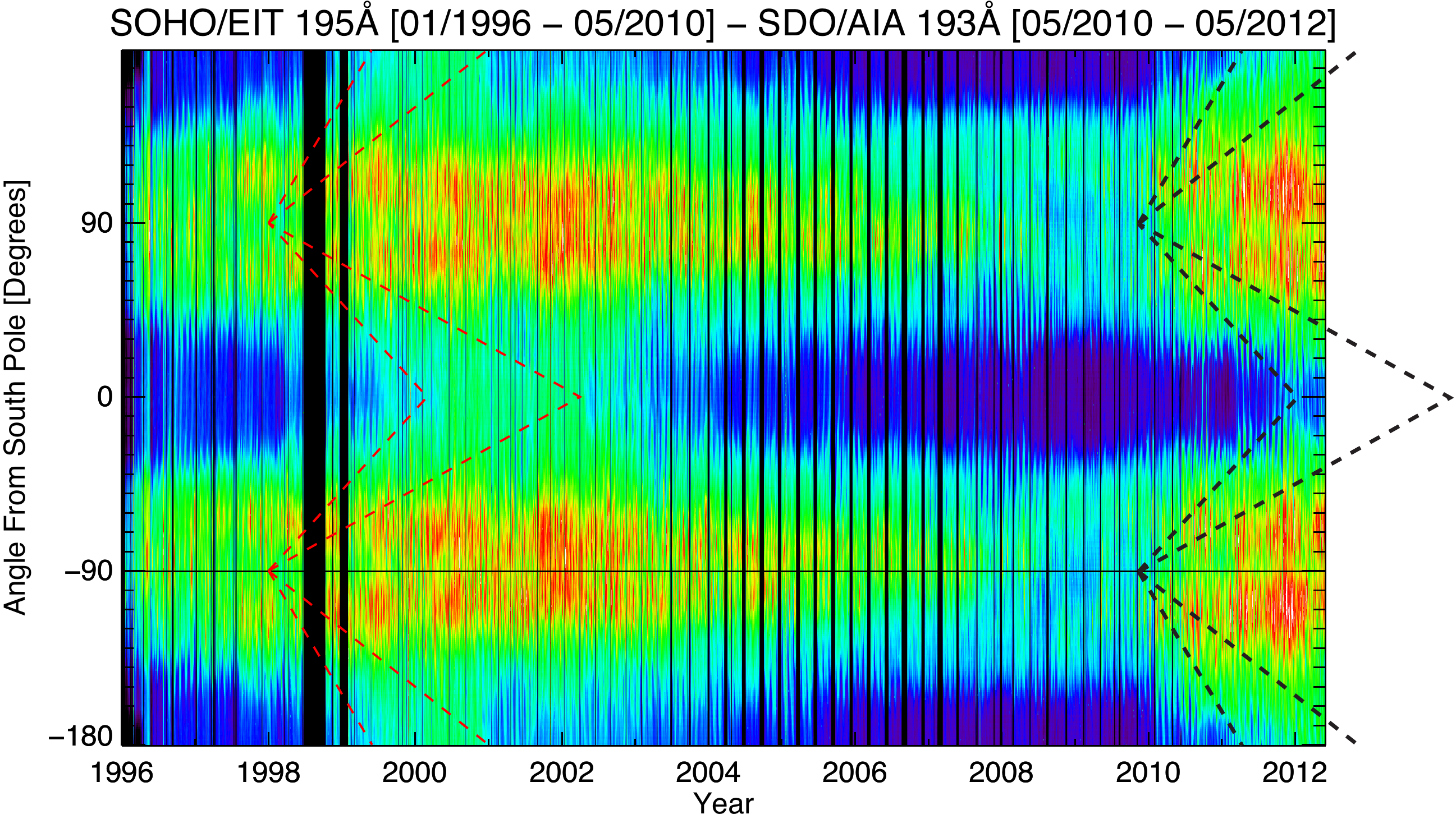}
\caption{Temporal evolution of the \ion{Fe}{12} coronal brightness integrated radially over a 1.15-1.25 R$_{\sun}$ annulus using \soho/EIT 195\AA{} and \sdo/AIA and 193\AA{} emission. This measure allows us to monitor the time-dependent behavior of the global coronal structure over solar cycle 23 and into the early part of cycle 24 as a function of the position angle of the emission from the south pole. For reference red-dashed guidelines are drawn from January~1, 1998 to highlight the different timing of the emission reaching and retreating from the poles in each hemisphere, the black dashed lines are duplicates translated to January~1, 2010 and are used to indicate changes between the ascending phase of cycles 23 and 24.\label{f7}} 
\end{figure}

Is there an observational means to explore the implied morphological/topological changes of the corona over the solar cycle? Adopting the approach of \citet{2011SoPh..tmp...24A} we can monitor the variation in integrated brightness of the \ion{Fe}{12} corona in a 1.15--1.25 R$_{\sun}$ annulus using the \soho/EIT 195\AA{} and \sdo/AIA 193\AA{} channels. The variation in the emission as a function of position angle identify specific trends related to the timing of the polar reversal and evolution of the global coronal structure in the declining phase of the cycle and into the early part of cycle 24. In Fig.~\pref{f7} we show the variation of the (exposure) normalized annular brightness of the EUV images as a function of position angle from the south pole (the north pole is at $-180$ and $+180$ degrees) from 1996 through the middle of 2012. A single daily image is used to isolate the annuli, they are then integrated over the radial distance to represent the average coronal intensity in \ion{Fe}{12} (about 1.5~MK) as a function of position angle. We note that much of the coronal emission is contained within about 55 degrees of the equator over the entire cycle with only poleward excursions at solar maximum.

Using the red dashed lines as a reference---drawn from an arbitrary starting point of Jan 1,~1998---we highlight the transport of coronal emission to high latitudes at solar maximum that is forced by the poleward advection of the magnetic field that is responsible for the polar magnetic reversal \citep[e.g.,][]{2011SoPh..tmp...24A, 2011ApJ...736..136W}, or the ``rush to the poles'' as it has been dubbed. Conceptually it is simple, the opposite polarity magnetic flux moving poleward will form helmet streamers in the corona that increase (or decrease in the south) in latitude as we progress towards solar maximum---by monitoring the emission in the extracted image annulus we should be able to identify these migratory time scales. Indeed, we see that the northern emission reaches the polar region faster (in 1999) and within eighteen months it is receding again, while the southern hemisphere has barely started its progression to the pole (2000). From this diagnostic plot, it takes at least a year longer before the high latitude southern emission begins to recede (early 2002). We also see the longer preponderance of emission in the southern hemisphere into the deepest part of minimum. It would appear that the asymmetry in the emission of the hemispheres is driven by, or responding to, offsets in their meridional transport and that the time difference at solar maximum has a prolonged effect lasting well into the declining phase of the cycle. 

It is also very interesting to note that, continuing the EIT record with AIA, we see that the northern coronal emission has reached the highest latitudes shortly before 2012 while the southern hemisphere is lagging far behind at the time of writing this manuscript (e.g., the black dashed linesÑduplicates of the red lines and translated in time). It is hard to predict when the southern coronal emission will reach the highest latitudes, but it seems to be at least a year out of phase with the north if you use the black dashed lines (drawn from Jan 1, 2010) as a visual guide. In this case the northern emission has reached the pole and the south seems to be much farther behind than this time last cycleÑhowever, predicting what will happen next is far beyond the scope of this article.

\begin{figure}[!ht]
\plotone{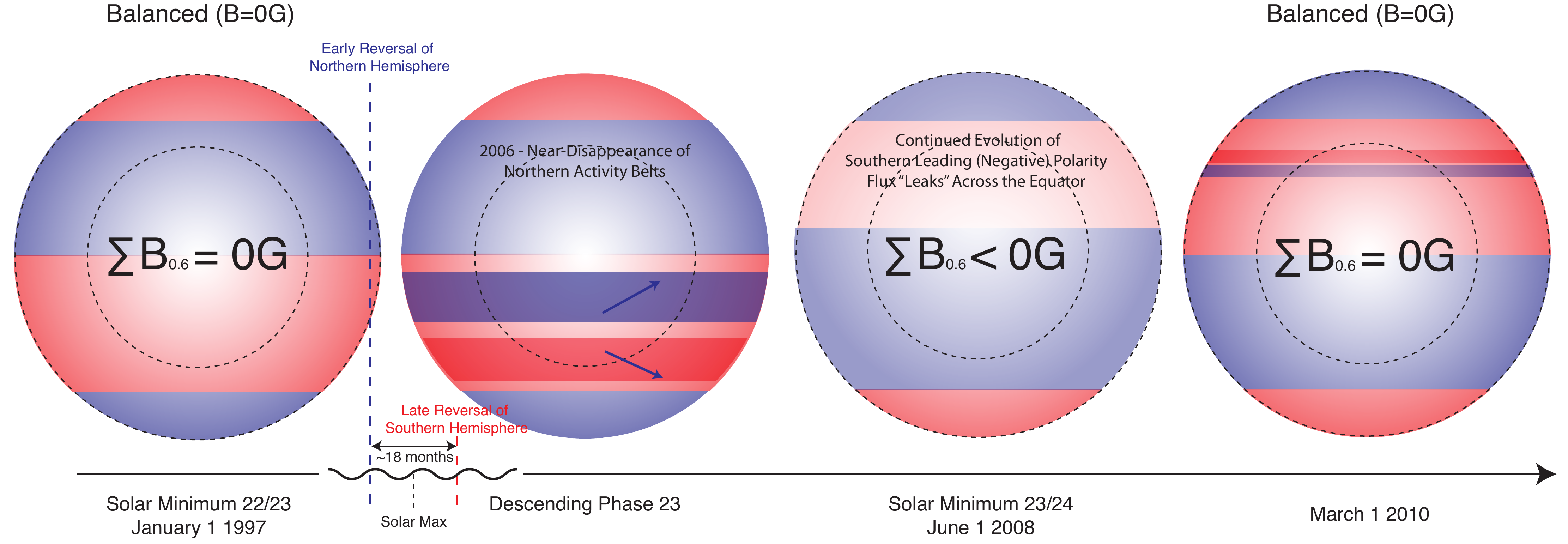}
\caption{A cartoon to illustrate the evolution of the underlying magnetism. Left, we show the 1997 solar minimum dipolar state (when the TSF of the equatorial region is approximately zero). This is followed by the early ascending phase of cycle 24 in 1998 when the new active latitude bands superimpose themselves. Those bands slowly migrate to lower latitudes and diffuse outwards from those towards the equator and polar regions to reverse the next cycle. Skipping forward the northern activity belt disappears in 2006 and progressing to the deep descending phase in mid-2008 where the TSF of the equatorial region is negative due in large part to the excursion of the southern negative band across the equator. Right, in early 2010, the disk appears to return to its dipolar state and the TSF has returned to zero. \label{f8}} 
\end{figure}

\subsection{A Possible Interpretation?}
The factors presented in the earlier subsections together would suggest that, following its late polar reversal, the evolution of the southern hemisphere proceeded at a slower rate (even though the northern cycle 24 field has begun to emerge in the equatorial region) and make its passage to the poles to perform the cycle 24 polarity reversal. In addition, the activity belt of the northern hemisphere seemed to disappear in late 2006 and there was a southern intrusion of negative polarity flux into the northern hemisphere. The components would naturally lead to the equatorial (and northern) portion of the MDI region being predominantly negative in sign while the southern portion is still relatively well mixed, i.e., it is ``typical'' quiet Sun. The sum of these effects would be very slightly, but persistently negative TSF, until the positively signed weak field driven by the early flux emergence in the northern hemisphere which combines to balance the equatorial region late in 2009.

To illustrate this physical picture we present Fig.~\pref{f8} where positive polarities are red and negative polarities in blue. The leftmost panel shows the situation in the cycle 22/23 minimum where the equatorial region TSF is zero (with a relatively large variance---see above) and the corona appears largely dipolar. Progressing through the ascending phase, the emergence of the cycle 23 activity belts continues, building to solar maximum in 2000-2002 where we recall that the northern hemispheric polar reversal (around 2000) was rapid and that the southern hemisphere reversed later and slower---lagging by the best part of 12 months. In the descending phase of the cycle the magnetic balance of the atmosphere is complicated by the northern activity belt disappearing in 2006 such that deep in the descending phase of cycle 23 (June 2008) the southern negative polarity flux band ventures across the equator and into the northern hemisphere. The slightly earlier appearance of activity in the northern hemisphere forces the flux balance of the equatorial region (TSF recovering to zero) and the dipolar corona of early 2010 (right).

\subsection{Cartoon Validation? Latitudinal Distribution of EUV Bright Points}
If our cartoon is even close to qualitatively correct we must consider a test, and one that is sensitive to the mixture of small scale magnetic polarities. One possible example lies in the number of EUV bright points (BPs) and their latitudinal distribution. BPs require mixed polarity field concentrations to be adjacent to one another \citep[most likely neighboring supergranular network vertices of opposing sign;][]{McIntosh2007} and so they can be used as a relative measure of the degree of mixed polarities present in a given region. 

\begin{figure}[!ht]
\plotone{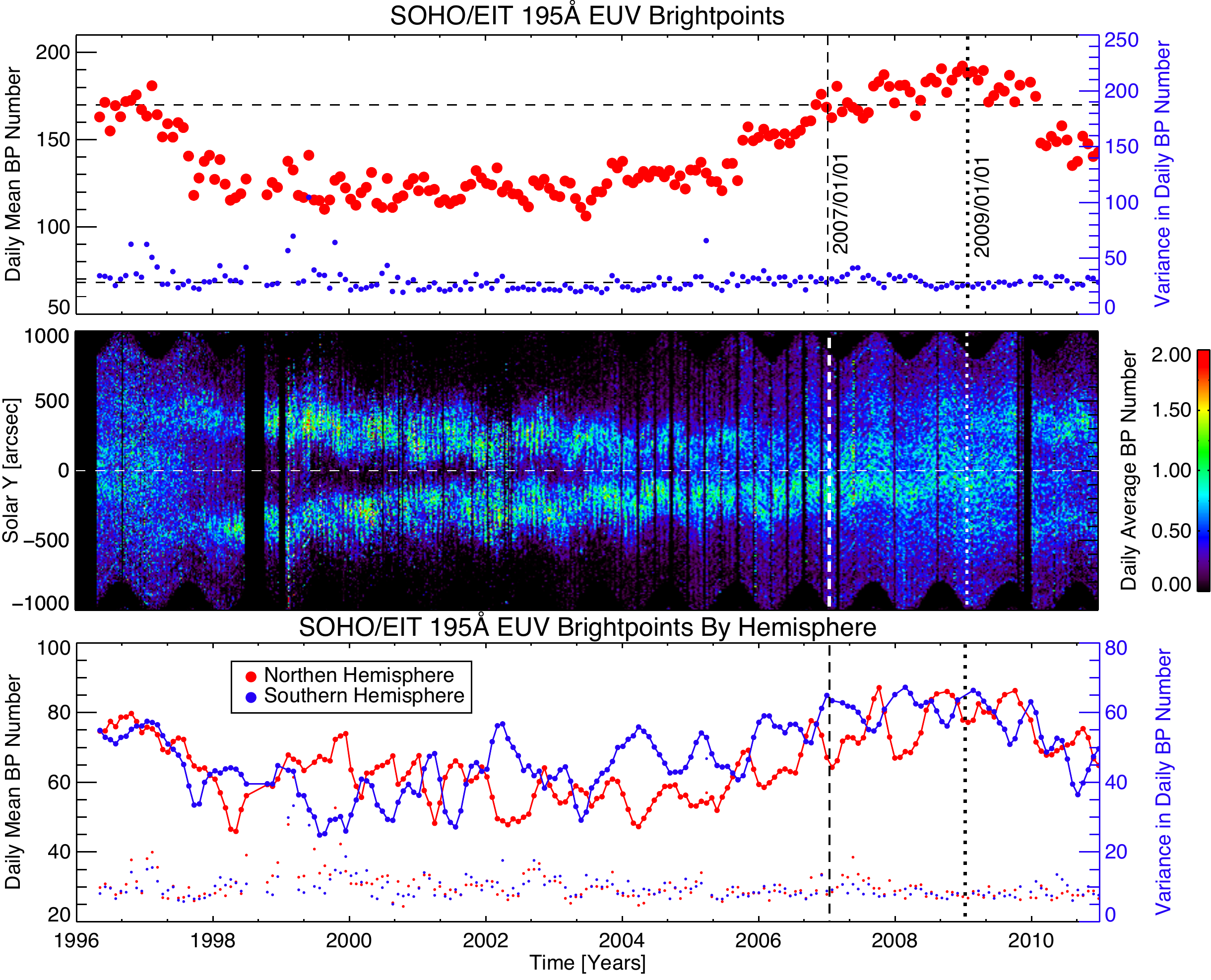}
\caption{Coronal EUV brightpoint (BP) evolution as observed in the \soho/EIT 195\AA{} passband over solar cycle 23 through the start of cycle 24. From top to bottom the panels of the figure show the daily average BP number and variance (red and blue dots respectively), the latitudinal distribution of BPs corrected for the solar B angle, and the daily BP number and their variances by hemisphere (red; north \-- blue; south). \label{f9}} 
\end{figure}

Figure~\pref{f9} presents initial results from the complete \soho/EIT 195\AA{} BP database \citep[see, e.g.,][and a forthcoming paper]{McIntosh2005}. The top panel shows the variation of the daily average BP number (red dots) and its variance (blue dots) over the solar cycle. We see that the 1996/1997 BP number is reached in 2007 and continues to increase upward before decreasing dramatically at the end of 2009. This is consistent with panels D and E of Fig.~\pref{f1} where the supergranular scale and MRoI power laws indicate smaller magnetic field length scales in 2009---smaller supergranules should lead to more BPs as the network vertices are systematically closer together \citep[][]{2011ApJ...730L...3M}. The middle panel of the figure shows the latitudinal variation of the BP distribution (corrected for the solar\--B angle). We see the equator-ward progression of BPs typical of a butterfly diagram, except that there is a systematically larger BP density in the southern hemisphere in the entire declining phase of the cycle---at this time the northern active latitudes have decayed considerably, as noted above. Through the descending phase there is a profound north-south asymmetry in BP density through 2009 when the BP density begins to grow again at high latitudes in the north. The bottom panel shows the balance between the north (red) and south (blue) BP distributions (for latitudes less than 60 degrees) and, as we have pointed out above, the two hemispheres seem to be well balanced with BPs through 2003 before the southern hemisphere appears to have systematically more BPs, a situation that persists until 2009 when the distributions appear to balance one another. 

Therefore, taking these measurements literally as a measure of locally mixed polarities the southern hemisphere would appear to have more mixed locally polarities while the northern hemisphere has less in the declining phase of the cycle. We believe that this simple measurement adds weight to the concept presented above concerning the temporal lag between the onset of the poleward flux migrations in each hemisphere, the ``disappearance'' of the northern activity belt, and the northward migration of the southern negative flux. Further, we see the strong correspondence between the BP number and the outputs of Fig.~\pref{f1} (\pref{f2}, and~\pref{f9}) that show an ``abnormal'' level of change during the extended solar minimum period period. Future investigations using the BP database (from SOHO and SDO) will use the pioneering work of \citet{Wilson1988} to study the apparently strong latitudinal overlap of activity cycles during the solar minima (2006-2010) and the patterns of meridional flow \citep[e.g.,][]{2009ApJ...702L..32S} among other things.

\section{A Broader Canvas: Historical Hemispheric Sunspot Numbers and Areas}\label{sunspot}
Based on the proceeding discussion we now look for signatures of hemispheric offsets in longer time records of solar activity. While the sunspot number is not a perfect record (in the sense that it does not capture the complete hemispheric magnetic flux distribution) any gross imbalance in the hemispheric magnetism of the Sun should be reflected in its hemispheric decomposition. 

\begin{figure}[!ht]
\plotone{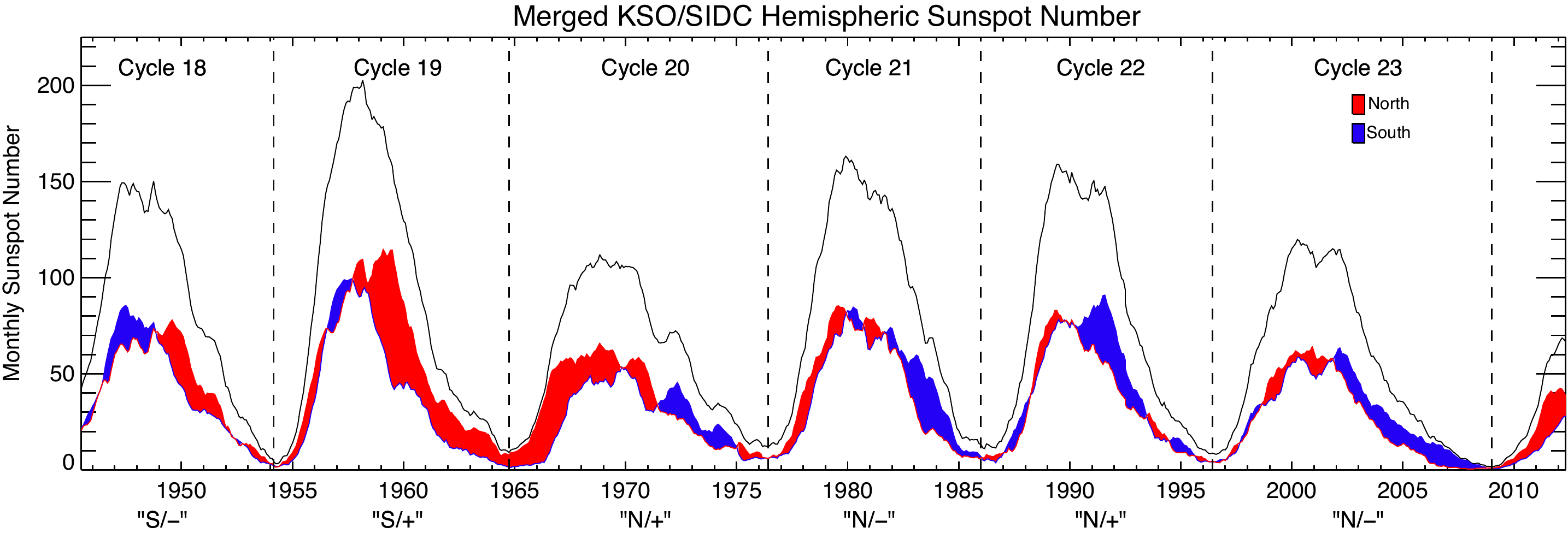}
\caption{The international sunspot number by hemisphere combined from data provided by KSO and ROB/SIDC \citep[e.g.,][]{2006A&A...447..735T} from 1945 to the present. The northern (red) and southern (blue) hemispheric sunspot numbers are shown against the total of the two (black). The shading of the difference between the northern and southern sunspot numbers indicates an excess in sunspot number between them. The dashed vertical lines are drawn from sunspot minima to delineate the cycles over the timeframe. The tags at the bottom of the figure indicate the hemisphere with most spots in the ascending phase of the cycle and the polarity of the magnetic field being advected into that hemisphere in the declining phase of the cycle. \label{f10}} 
\end{figure}

\begin{figure}[!ht]
\plotone{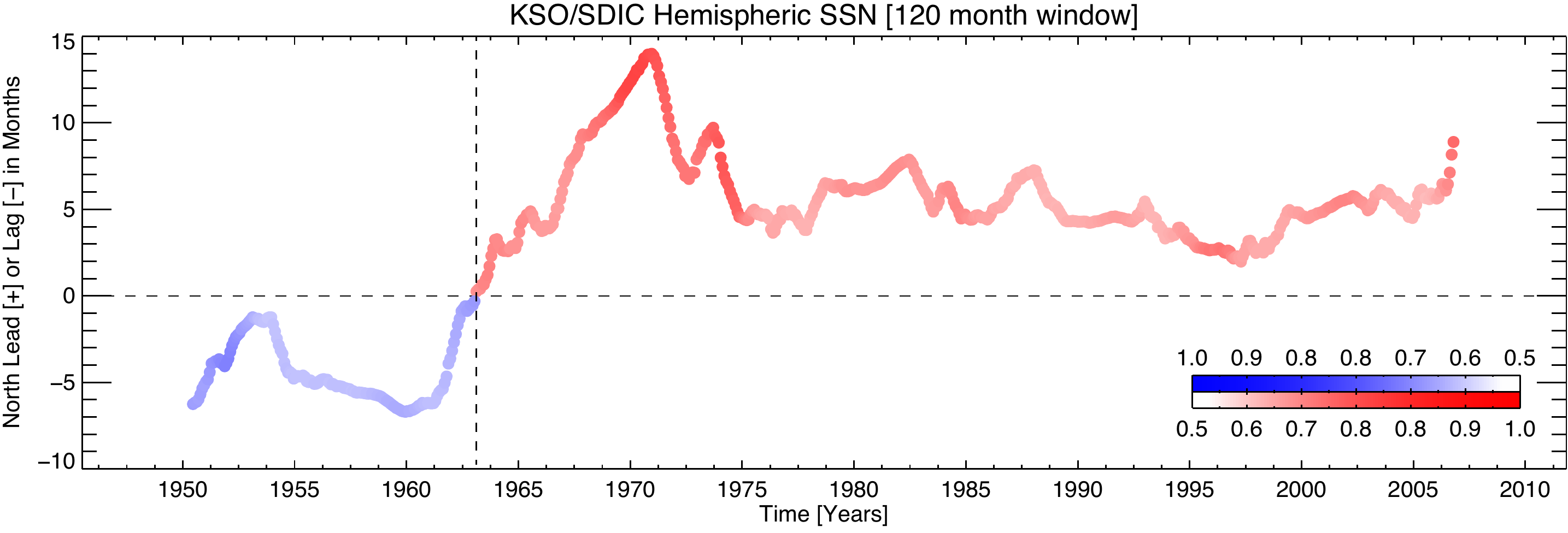}
\caption{Using a 120\--month long averaging window we measure the lag time of the hemispheric sunspot numbers shown in Fig.~\pref{f10}. Increasing color depth indicates the magnitude of the cross-correlation coefficient while the color itself dictates whether the north (red) or south (blue) is leading. \label{f11}} 
\end{figure}

This hemispheric ``tug-of-war'' can be better appreciated by considering Fig.~\pref{f10} where we plot the combined sunspot record of the Kanzelh\"{o}he Solar Observatory (KSO) and the Solar Influences Data Center (SIDC) housed at the Royal Observatory of Brussels (ROB). The KSO and ROB/SIDC took great care to count the number of northern and southern sunspots separately \citep[][]{2006A&A...447..735T} and we show the total sunspot number (black solid line) as well as the number counted in the northern (red) and southern (blue) hemispheres. The vertical dashed lines denote the sunspot minima and the tags underlying the figure indicate the leading hemisphere of the ascending phase and the sign of the earlier pole-canceling magnetic field. The shading illustrates the temporal variation of hemispheric ``dominance'' (which hemisphere has the largest sunspot number) and we can see that a pattern of the northern hemisphere leading solar activity started around 1965, as implied by \citet{2006GeoRL..3303105H}, where something quite unusual must have happened over the course of cycle 19 as this seems to be the point where the north ``overtook'' the south \citep[as also discussed by][]{2006GeoRL..3303105H}. Figure~\pref{f11} uses the cross-correlation of a 120\--month wide moving window of the northern and southern sunspot number timeseries to provide a lower estimate for the time lag between the hemispheres. We see that since 1963 the magnetic activity of the northern hemisphere has been leading by at least five months.

\begin{figure}[!ht] 
\plotone{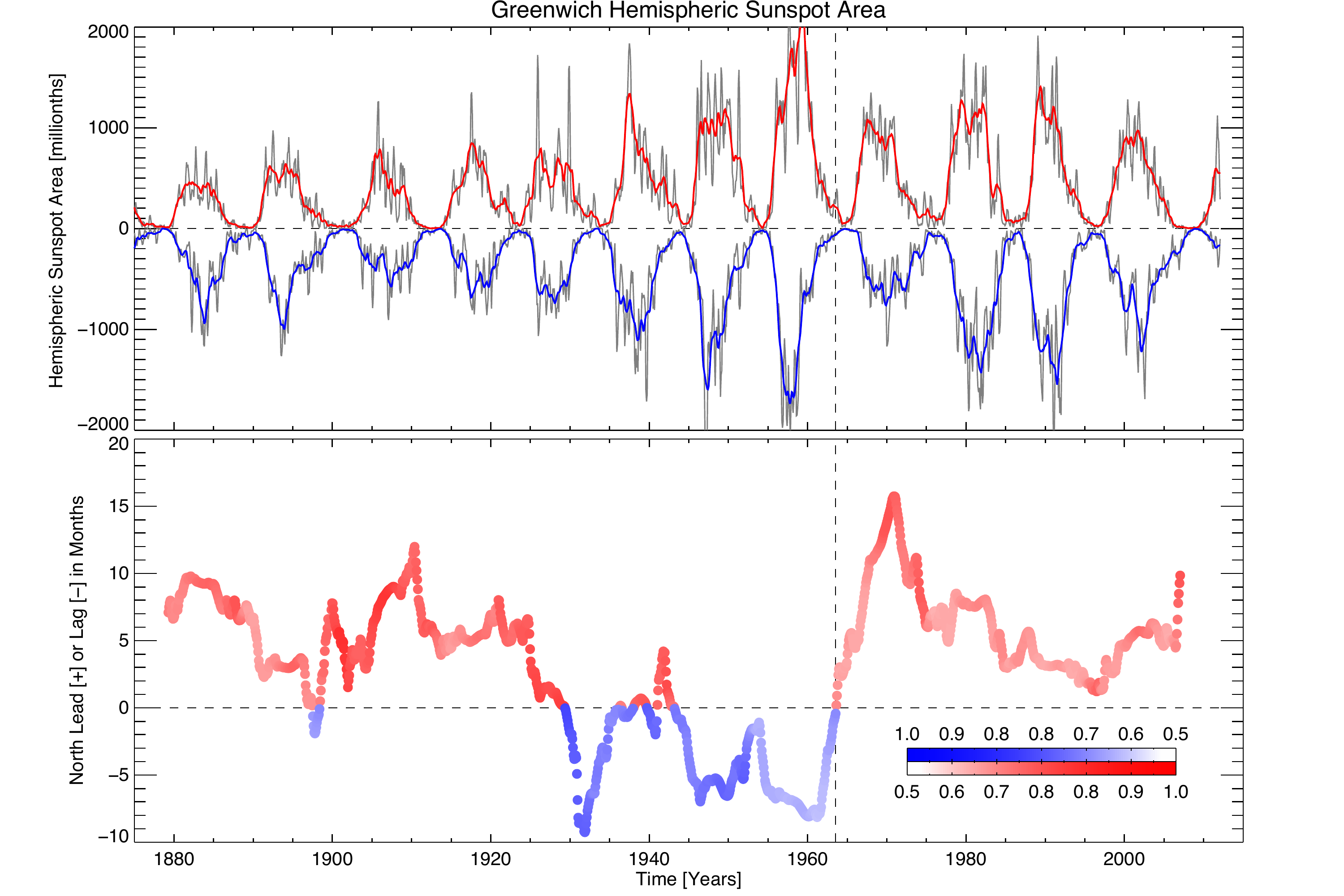}
\caption{Top and bottom we show the Greenwich sunspot areas differentiated by hemisphere and the 120 month averaged cross-correlation lag. Increasing color depth indicates the magnitude of the cross-correlation coefficient while the color itself dictates whether the north (red) or south (blue) is leading. \label{f12}} 
\end{figure}

Only a few observational records extend further back in time that are differentiated between activity in the northern and southern hemispheres. One such record is the Greenwich sunspot area\footnote{\url{http://solarscience.msfc.nasa.gov/greenwch.shtml}} where the monthly averaged northern (red) and southern (blue) values (measured in millionths of the disk area) are shown in Fig.~\ref{f12}. The bottom panel again uses a 120\--month window to estimate the lag time of the hemispheric sunspot areas. We see that the post-1950 variation matches almost perfectly with that of Fig.~\pref{f11} and note the southern lead back to 1928 following another spell of northern dominance. So, over the 130 years of the Greenwich record, the dominant hemisphere has changed twice. For the sake of comparison, studies of the Ca IIK image archive conducted by \citet{2011ApJ...730...51S} show additional evidence of activity asymmetries in the historical observation record (their Figs.~4 and~7 for north and south leading cases, respectively).

Our brief digression into the historical activity records illustrates that hemispheric imbalances and apparent lags in activity are a prevalent behavior and are {\em not} anomalous. What may be anomalous are the effects that these imbalances have on small (length) scale energy release processes (e.g., Fig.~\pref{f1}) and the global length scales, i.e., those of the heliosphere (e.g., Fig.~\pref{f2}). In the following sections we will look at the possible impacts that prolonged hemispheric asymmetries may have on each of these scales, starting with the larger of the two.

\section{Impact on the Heliospheric Scale?: The Large 2009 Cosmic Ray Flux}
In Fig.~\pref{f2} we saw the unexpected rise of the CRF at the depths of the last solar minimum. We deduce that, based on the incredible temporal correspondence, the underlying solar magnetism contributed to that rise. While the Earth's magnetic field deflects some of the CRF affecting the Earth, it is the shape and strength of the Sun's magnetic field and plasma outflow (the solar wind) that forms the solar system's primary ``deflector shield,'' the heliosphere, which takes the brunt of the particle bombardment. Indeed, it is the latter, and its (approximate) 11-year variation, that is responsible for the cyclic modulation of the CRF measured at Earth \citep[e.g., Fig.~\pref{f2} and][]{2005JGRA..11012108U}.

\begin{figure}[!ht] 
\plotone{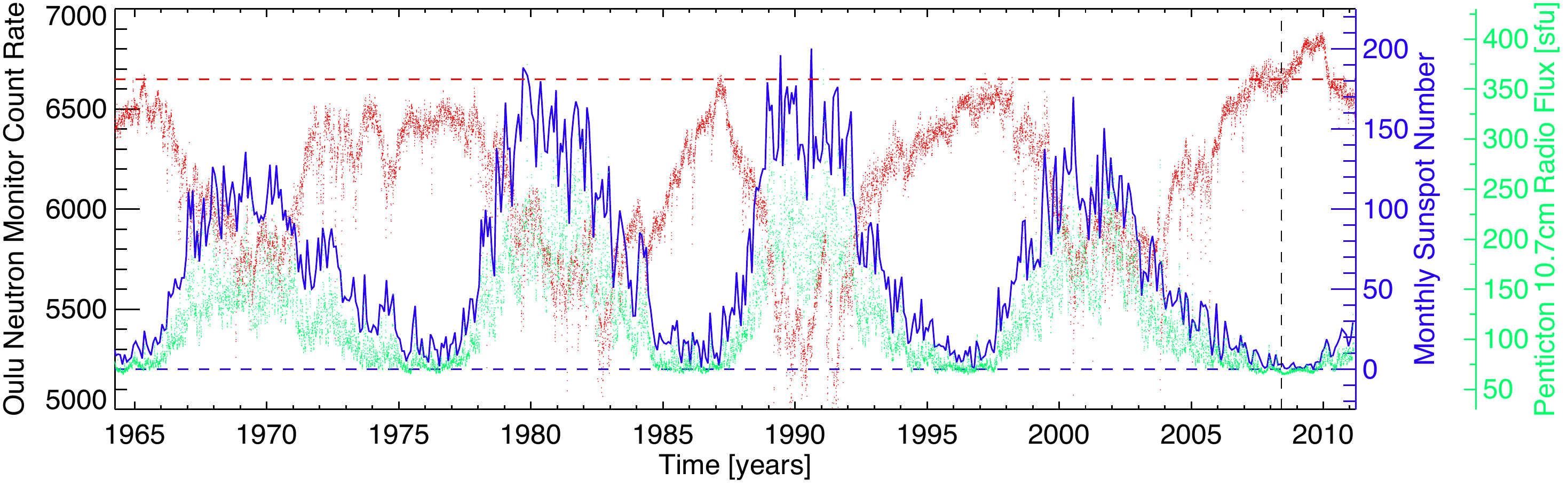}
\caption{The CRF as measured by the neutron monitor at the Oulu Station of the Sodankyla Geophysical Observatory in Finland (red), versus the monthly sunspot number (blue) and Penticton 10.7cm radio flux (green) proxies of solar activity. Notice the approximate 11-year periodicity in the sunspot number and the anti-correlation with the measured CRF. Notice also the enhanced solar-minimum CRF starting in early 2008 and its rapid decrease in early 2010. The horizontal red line marks the highest, pre-1997 CRF (measured in the spring of 1965) that is 4\% lower than the levels measured in mid 2009. The dashed vertical black line marks (2008 June~1) and indicates the start of the anomalous rise in the CRF during the most recent solar minimum. \label{f13}} 
\end{figure}

In Fig.~\pref{f13} we clearly see the cyclic variation in the sunspot number (blue) and 10.7cm radio flux (green), commonly accepted proxies of solar surface magnetism, and the CRF (red) measured at the Oulu station in northern Finland from the early 1960s to earlier this year. Over the last five solar activity cycles, the CRF has alternated between flat and peaked maxima driven by the efficiency with which the positively charged particles can penetrate into the inner heliosphere when the heliospheric field is parallel (peaked) or anti-parallel (flat) to the galactic field---when drifting inward on a predominantly inwardly pointing (negative) solar magnetic field the particles suffer less scattering and so the CRF that reaches Earth is increased \citep{2005JGRA..11012108U}. 

Strikingly, the anomalously high CRF of the recent solar minimum (2008--2009) is 4\% larger than ever before recorded. In the lower energy bands detectable from space the increase in the CRF reached levels of greater than 20\% \citep{2010ApJ...723L...1M}, and this anomalous increase is reproduced at all neutron monitor stations: is not isolated to those near the Earth's polar regions. Further, from the \soho/SSR measurements shown in Fig.~\pref{f2}, it is clear the increase in the CRF is not driven by processes nearer to the Earth than \soho.

\begin{figure}[!ht] 
\plotone{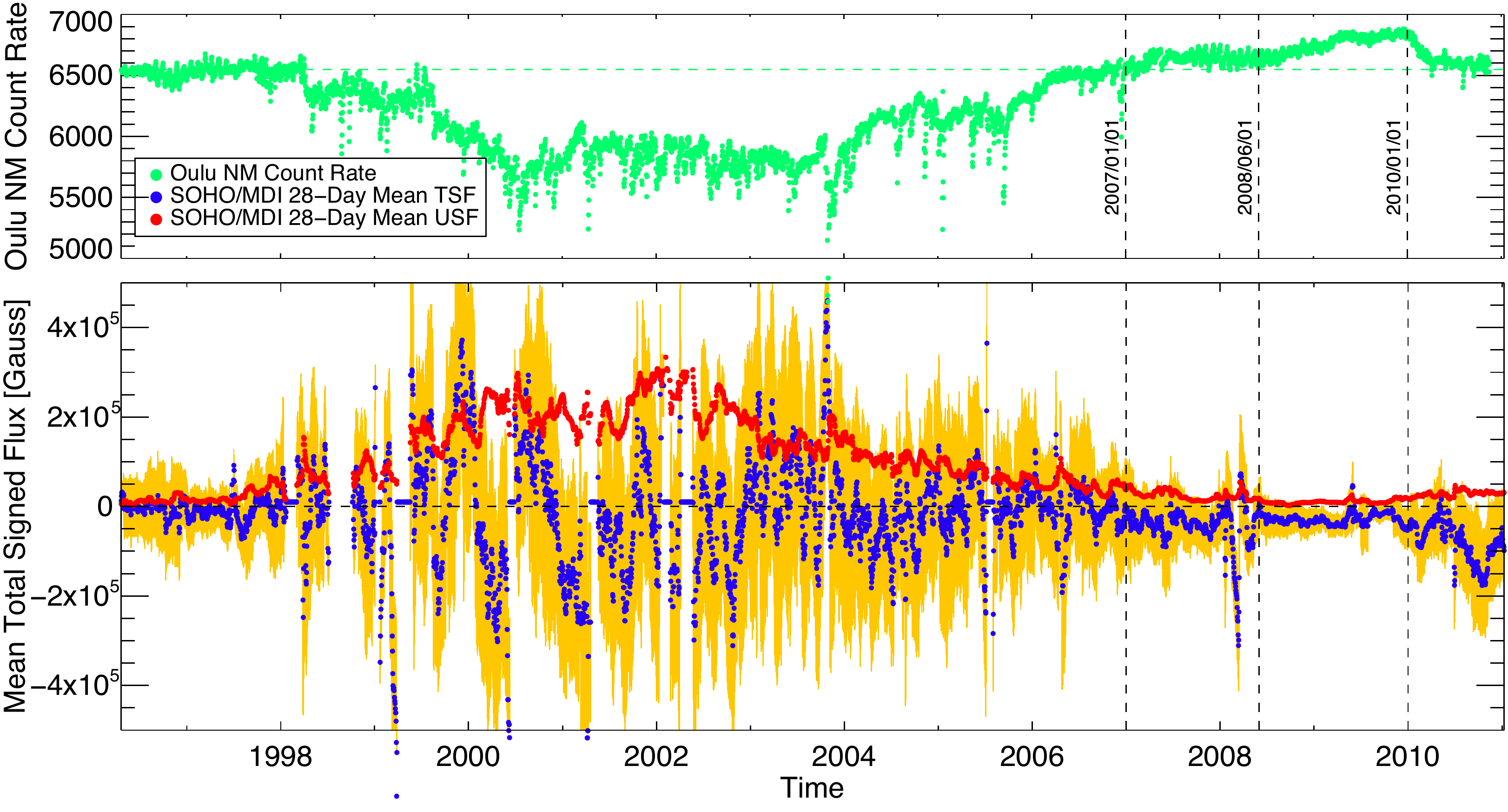}
\caption{The \soho{} era CRF as measured by the neutron monitor at the Oulu Station of the Sodankyla Geophysical Observatory in Finland (green; Fig.~\pref{f2}) compared with the equatorial \soho/MDI magnetic field metrics (Fig.~\pref{f4}). \label{f14}} 
\end{figure}

We see that, comparing the CRF and TSF curves of Fig.~\pref{f14}, the period of marking the start of the equatorial magnetic field imbalance coincides in time with the enhancement of the CRF over that of the 1996/1997 solar minimum. More significantly, the period when the CRF reaches its all-time high value is when the magnetic field measured around the center of the Solar disk is predominantly negative and shows very little variance. So, not only is there an imbalance of one polarity over the other, but the field strengths themselves are very low. The rapid turnover in the CRF (about 2010 January~1) occurs at the time when the variance of the magnetic field again increases and the TSF begins to migrate back towards zero, i.e., when the imbalance has been redressed.

Given the rapid response of the CRF to changes in solar magnetic field (App.~\pref{oscill}), we deduce that the apparent lag of the northern and southern solar hemispheric activity in the declining phase of solar cycle 23 and into the 2009 minimum caused a lengthy asymmetry to develop in the structure of the heliosphere. For instance, consider the previously noted work where one pole (say the north) takes 12 months longer than the other pole to reverse sign. The natural consequence of that will be that, for a portion of that 12 months, a large portion of the SunÕs surface magnetic field will be of the opposite sign and will naturally give rise to more regions of open magnetic field in the declining phase. 
In the central panel of Fig.~\pref{f5} notice how far north the open negative field (green traced field lines) have encroached into the northern hemisphere. We stress that such a magnetic imbalance would also naturally lead to a solar corona, and inner heliosphere, that is punctuated with equatorial corona holes \citep[e.g.,][]{2006ApJ...644L..87M}---a staple of the descending phase of most solar cycles \citep[e.g.,][]{1981JGR....86.2079H,1998SoPh..177..375F}. These coronal holes will be of mixed size (depending on the local magnetic field distribution) and a heliosphere that is not dipolar, lacks internal rigidity (with a mental image of ``Swiss cheese''). We speculate that this lack of organized global closed structures in the heliosphere, driven by the hemispheric asymmetry of the surface magnetism, plays an important role in permitting more cosmic rays into the near-Earth system \citep{2010ApJ...723L...1M}. 

Finally, we note the strong anti-correlation of the CRF in the ascending phase of cycles and the apparently cycle dependent lag of the CRF from the variation in the sunspot number. It would appear that, in the descending phase of solar cycles the CRF measured at Earth increases at a rate shallower than the drop off of the Sunspot number---an issue we will return to below.

\section{Impact on the Small-Scale: Energetic and Particulate Output}\label{sscale}
%

Comparing the panels of Fig.~\pref{f1} (and Figs.~\pref{f6},~\pref{f7}, and~\pref{f9}) with Fig.~\pref{f4} (and~Fig.~\pref{f14}) we deduce that a strong relationship exist between the negative TSF, weak USF (and small variance) in the extended phase of the recent solar minimum (June 1 2008 -- Jan 1 2010) and the reduced radiative and particulate output of our star. That extended ``depth'' of the solar minimum was coincident in time with a reduction in the apparent size of the quiet magnetic network and that, we propose, caused many of the other energetic manifestations with the likely exception of the increased CRF. We still support the conclusion of \citep{2011ApJ...730L...3M} \citep[and][]{2011ApJ...740L..23M} that the reduced energetics of the corona and solar wind during the solar minimum were driven by the reduction in ``complexity'' of the supergranular network vertices that are the mass tributaries into the open and closed outer atmosphere alike \citep[e.g.,][]{Hassler1999,McIntosh2007}. It remains to be seen how the complexity, or discrete flux element filling factor of the network vertices, moderates the energy and mass flow into the outer atmosphere but those (necessary) high-resolution detailed studies are beyond the scope of this Paper. 

\begin{figure}[!ht]
\plotone{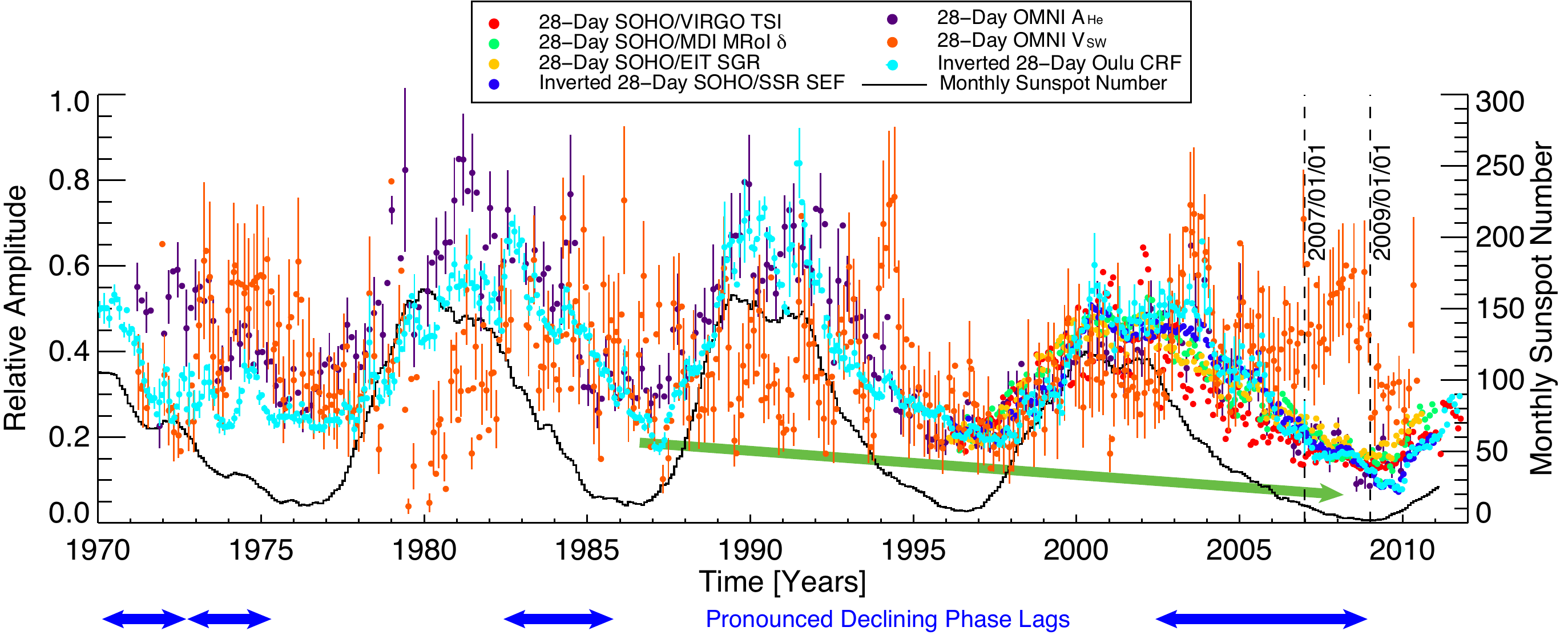}
\caption{Comparing the relative amplitude of the small and global spatial scale measurements used in the figures above from the present day back to 1970. We show the 28-day averaged values of the \soho/VIRGO TSI (red dots); \soho/MDI MRoI $\delta$ (green dots); \soho/EIT Supergranular Network Radius $\delta$ (yellow dots); the ``inverted'' \soho/SSR Single-Event-Flag Count (dark blue dots) with the longer solar wind speed (\vsw; orange dots) Helium abundance (\ahe; purple dots) from the OMNI database, with the ``inverted'' Oulu CRF (turquoise dots). Vertical dashes indicate the variance in that measurement over the temporal averaging applied. All of the timeseries (with the exception of \vsw) are scaled to have the same dynamic range between 2001 and 2008 so that their relative amplitude changes can be compared and put in contrast with the longer records. The term ``inverted'' refers to a timeseries that has been rotated around its horizontal axis such that a peak is now a trough. See the online edition of the Journal for a supporting animation of this superposition of different timeseries. \label{f15}} 
\end{figure}

Out of curiosity we construct Fig.~\pref{f15}. The figure (and accompanying animation that is available in the online edition of this Journal) shows the timeseries shown in earlier figures (see caption for details) that are now rescaled in amplitude such that they have the same dynamic range between 2001 and 2008 (with the exception of the OMNI solar wind speed \vsw; orange dots). The \soho/SSR SEF and Oulu CRF timeseries have been ``inverted''---reflected in the horizontal plane such that a peak in the original timeseries is a trough in the inverted and vice versa. Several things are immediately visible:
\begin{enumerate}
\item{The timeseries appear to be strongly in phase with the sunspot number in the ascending phase of all cycles,} 
\item{As noted above, the timeseries have a time varying lag (or temporal offset) with the sunspot number in the declining phase of several cycles with cycles 21 and 23 showing pronounced lags,}
\item{\vsw{} doesn't appear to vary at all in phase with the sunspot cycle although we notice that periods of sustained high average speed occur when hemispheric asymmetries are still present very late in the cycle (e.g., 1975 and 2008), and}
\item{As noticed by \citet{2011ApJ...740L..23M}, the downward trend of the quantities most directly linked to the distribution and magnitude of surface magnetism between 1985 and 2010 is clearly visible.}
\item{Finally, it appears that the small-scale measures of radiative and particular output from our star are related to the global-scale structure of the heliosphere as manifested by its effectiveness to repel cosmic rays.}
\end{enumerate}
As demonstrated in the supporting animation of the figure we see that the similarity of the timeseries in this rescaled format is truly remarkable---the supporting online animation of this figure emphasizes the incredible agreement. Note that we attribute the lack of a strong correspondence between the \vsw{} and the sunspot number to at least two things; the measurement of one point in 3-dimensional space being a characteristic of the global behavior, except at times when there is a globally coherent open flux pattern (like that of 2009). It could also be that the value of \vsw{} at 1AU is not directly related to surface magnetism and is more likely a property of wave dissipation processes taking place en route to the measuring spacecraft \citep[][]{2011Natur.475..477M} except again, when the global field is relatively well structured over a long time.

\section{Discussion}
An asymmetry in the Sun's low-latitude magnetic field persisted for
nearly three years, including an 18-month spell of very weak,
imbalanced, field. This weak imbalanced field period coinciding with an
unexpected decrease in the radiative and particulate output of our star
as well as an increase of the cosmic ray flux bathing the Earth. In the
preceding sections we have deduced that this prolonged period of
imbalance is the result of subtle changes of the magnetism in the
northern and southern hemispheres during the descending phase of the
last solar cycle. Furthermore, we have demonstrated that hemispheric
asymmetries of the photospheric magnetic field are the norm and not
necessarily unusual. These driven changes in the photospheric magnetic
field distribution have a direct effect on the radiative and particulate
output of our star although we are far from understanding the detailed
mass and energy transport processes at the heart of the
radiative/ particulate output. It follows that the asymmetry of the
photospheric magnetic field produces an asymmetric heliospheric magnetic
field and the modulation of measures like the CRF (intrinsically tied to
the larger scale evolution of the heliospheric magnetic field) vary with
remarkable correspondence to those tied to much smaller spatial scales
(Fig.~\pref{f15}). This coupled behavior would suggest that system
modeling of the heliosphere is necessary to accurately forecast future
behavior. Such forecasts should stem from a better observational
understanding of the physical processes which govern the detailed
distribution and balance of surface magnetism.

While the underlying physical cause in the convection zone (or below) is not known \citep[][]{2007AdSpR..40..899C}, our observational results would appear to the period of apparent magnetic imbalance seems to be forced by the hemispheric differences in the meridional circulation (the ``conveyor'' that recirculates the Sun's magnetic field under the surface) and magnetic diffusion rates \citep[e.g.,][]{2010Sci...327.1350H,2011Natur.471...80N}. Based on the observations presented above we can infer the following:
\begin{enumerate}
\item{The proximate cause of the negative TSF at disk center in the cycle 23/24 solar minimum is the greater activity in the southern hemisphere during the descending phase of cycle 23 (from 2003-2008). From the perspective of the mean poloidal field, the emergence of flux in an activity band in the ascending phase of cycles gives rise to a double-banded structure at high latitudes (see e.g., Fig~\pref{f8}). The equator-ward edge of each band diffuses toward the equator and reconnects across the equator with the corresponding flux in the opposite hemisphere. If there is more emergence in the southern hemisphere, only some of this trailing flux annihilates. This progression naturally gives rise to the imbalance and negative TSF at disk center at solar minimum.}
\item{It appears that the greater activity in the southern hemisphere in the descending phase of cycle 23 (2003-2008) is the result of a phase shift between the magnetic activity of the northern and southern hemispheres, as illustrated in Fig.~\pref{f10}.}
\item{We believe that the reason for this phase shift is the slower poleward transport and delayed reversal in the southern hemisphere that is illustrated in Fig~\pref{f7}. We attribute this delay to an asymmetry in the meridional flow between 1996-2003 \citep[e.g.,][]{2010Sci...327.1350H,2010ApJ...717..488B}.}
\item{We don't really know why the meridional flow was asymmetric. One possibility is that an asymmetric flow can be attributed to random fluctuations of the motion; as a weak flow component, the meridional flow commonly undergoes chaotic fluctuations in global convection models which would (plausibly) average out over long times, so shifting phases as seen in Fig.~\pref{f10} would be expected. We note that it only requires an impulsive variation in the meridional flow pattern to throw the phase of activity off---this does not require a persistent asymmetry---that would cause the phase shift to slowly increase steadily with time, which it does not appear to do. Fig.~\pref{f11} suggests that impulsive, random shifts (in either the meridional flow or the timing of flux emergence or both) occur, but those are coupled with a dynamo relaxation time of 4 cycles where the long-term coupling acts to ``lock'' the phase of the hemispheres.}
\end{enumerate}

We have seen that this phase of hemispheric asymmetry (and apparent reduction in strength) of the Sun's surface magnetism during the recent solar minimum had a direct impact on the structure of the small and large scale energetics of the outer solar atmosphere and heliosphere. For a considerable time their signature has been inferred from interplanetary space measurements of north-south variations in the CRF measured by {\em Ulysses} just before the cycle 22/23 minimum \citep[e.g.,][]{1996ApJ...465L..69S} to deviations in the (anticipated) sector structure of the solar wind \citep[e.g.,][]{1997JGR...102.4673C}. In an effort to explain the {\em Ulysses} measurements \citet{2000ApJ...533.1084S} coined the phrase ``ballerina skirt" to describe the shape and tilt of the Heliospheric Current Sheet (HCS). They illustrated that the variations in the net imbalance of the magnetic pressure in the northern and southern hemispheres in the declining phase of solar cycles (modulated also somewhat by the ram pressure of the wind) can affect the modesty of the ballerina, causing the Sun to push its skirt southward \citep[e.g.,][]{2001GeoRL..28...95M, 2003GeoRL..30vSSC2M, 2004SoPh..221..337M, 2007AdSpR..40.1034M}. The latter series of papers, motivated by the inference of \citet{1998GeoRL..25.1851M}, that significant departures from heliospheric magnetic symmetry could be inferred in the late-declining phase of many solar cycles from systematic changes in the sector structure of the annual variation in the solar wind speed. \citet{2003GeoRL..30vSSC2M} and \citet{2006GeoRL..3303105H} went further, identifying that the southward tilt of the ballerina's skirt has been a persistent feature of the late-declining phase of solar cycles going back into the late 1960s. In each of these cases one might expect that, based on the discussion of above, the magnetic evolution of the northern hemisphere was ahead of the south (by an appreciable amount of time) such that the early advection of the pole-canceling magnetic field gave rise to an north-south magnetic pressure imbalance that pushed the HCS southward in the late-declining phase of several recent solar cycles. Interestingly, the often reported (and confusing) anti-inclination and oppositely directed motion of the streamer belt and the HCS in the declining phase of cycles \citep[e.g.,][]{1997JGR...102.4673C}, is related to the former being driven by the pole-canceling magnetic field distribution. Indeed, it is precisely this streamer motion that is captured in the diagnostic of Fig.~\pref{f7} and used as one of the alternative measures of the magnetic evolution at the heart of this puzzle. It remains to be seen if this pattern of northern dominance will change soon enough for this to be the ballerina's last dance \citep[][]{2011A&A...525L..12M} unless the background solar magnetic field continues to weaken---as may be inferred from the analysis presented in the previous section \citep[and also in][]{2010arXiv1009.0784P}. 

Unfortunately the record of detailed magnetograms is short and other magnetogram records are rare such that we are limited to exploring the Sun's historical connection to the heliospheric-scale structure using model-dependent \citep{2004Natur.431.1084S} or chemical \citep{2005JGRA..11012108U, 2006SSRv..127..327S} proxy inference. Therefore, it becomes difficult for us to accurately assess the impact that these subtle changes of solar surface magnetism and the Sun's dynamo \citep[e.g.,][]{1955ApJ...122..293P,1999ApJ...518..508D,2005LRSP....2....2C,2010LRSP....7....1H} have on the Earth system on timescales commensurate with those affecting cloud production \citep{1997JASTP..59.1225S,2007A&G....48a..18S,2009GeoRL..3602803S,2010GeoRL..3703802C}, or the broader Sun-climate connection \citep[e.g.,][]{Kumala2010}.

\section{Conclusion}\label{conclusion}
%
 
The summer of 2009 saw a considerable low in the radiative output of the Sun that was temporally coincident with the largest cosmic ray flux ever measured at 1AU. Combining measurements and observations made by the \soho{} (and \sdo) spacecraft we have begun to explore the complexities of the descending phase of solar cycles. A hemispheric asymmetry in magnetic activity is observed and its evolution monitored. Studying historical sunspot records with this picture in mind shows that the northern hemisphere has been leading since the middle of the last century and that the hemispheric ``dominance'' has changed hands twice in the past 130 years. We have deduced, largely due to the very strong temporal correspondence, that the prolonged magnetic imbalance has a considerable impact on the structure and energetics of the heliosphere. While we cannot uniquely tie the variance and scale of the surface magnetism to the dwindling radiative and particulate output of the star, or the increased cosmic ray flux, through the 2009 minimum, the timing of the decline and rapid recovery in early 2010 would appear to inextricably link them. 

The observations presented above lend support to a picture where the hemispheres of the Sun {\em must} be studied and treated independently, and that the globally averaging of quantities minimizes our ability to consider the degrees of freedom in the coupled solar atmosphere. Furthermore, it appears that the hemispheres are appreciably out of phase with each other at the time of writing this manuscript and the basal energetics of that system are in clear decline. The observations presented give cause for concern, especially with respect to our present understanding of the processes that produce the surface magnetism in the (hidden) solar interior. However, we can state with some certainty that if the observed asymmetry in surface magnetism continues in-step with the slow decline in magnetism, then we fear for the modesty of aforementioned ballerina, and can anticipate that the CRF during the next solar minimum will exceed the values of 2009 while the radiative and particulate outputs reach previously unanticipated lows.
  
\acknowledgements
We are grateful to comments on earlier drafts of this manuscript by Eric Priest, Leon Golub, Hui Tian and SWM thanks Ineke De Moortel for her encouragement. The Oulu Neutron Monitor data were obtained from the Sodankyla Geophysical Observatory via the website \url{http://cosmicrays.oulu.fi/}. The research conducted by SWM was partly supported by National Science Foundation (NSF) grant ATM-0925177. SWM and RJL were also supported by NASA LWS grants NNX08AU30G and NNH08CC02C. LS, RSM, and SWM were also partly supported by NASA grant NNX08AL23G to produce the EIT BP database. 
Special thanks to Jeneen Sommers (Stanford) for providing MDI data on request. SOHO is a project of international collaboration between ESA and NASA to study the Sun. We are thankful for the online publication of the KSO and SIDC international sunspot numbers. The National Center for Atmospheric Research is sponsored by the National Science Foundation.

\appendix

\section{The SOHO Solid State Recorder as a Cosmic Ray Detector}\label{SSU}
SOHO's orbit is outside of the Earth's magnetosphere and so it is perfect for our demonstration of an extra-terrestrial source of the CRF modulation. The Solid State Recorder (SSR) on board SOHO is used as the primary mass storage for the spacecraft when not in direct contact with Earth and has a capacity of 2Gb. The SSR is protected against data corruption caused by energetic particles originating at the Sun by employing an error correction Hamming code developed by Astrium (the prime contractor of SOHO spacecraft for ESA). With such a code any single bit error in one word of data can be corrected. To properly reduce the bit error rate of the scientific data telemetered to ground stations the SSR data is continuously ``scrubbed'' to keep the memory clean---reading the data stored in memory and correcting bad bits as needed. When bad bits are detected and corrected the spacecraft management system increments a counter. Because SOHO is in a ``halo'' orbit at the Lagrange ``L1'' point between the Sun and the Earth, and thus outside the Earth's magnetosphere, this serendipitous error count is a measure of the energetic particle environment in the solar system not affected by space-storm activity driven near the Earth.

\section{The Cosmic Ray Flux}

The Earth is bathed in a flux of energetic particles originating from the Sun and outside of our heliosphere, called cosmic rays. While the smallest component of that cosmic ray flux (CRF), also consisting of the lowest energy particles, originates in energetic solar phenomena such as flares and coronal mass ejections, the majority arise in cataclysmic events outside the heliosphere that penetrate the solar system and eventually reach Earth \citep{2001AIPC..558...27G}. Indeed, cosmic rays constitute a non-negligible fraction of the annual radiation dosage of humans \citep{2010SpWea...800E04S}, pose a considerable threat to the well-being of astronauts during space flights \citep[e.g.,][]{2005ICRC....2..433M}, and have an impact on aircraft avionics and airborne computer systems \citep{1993ITNS...40..120T,Ziegler1996}. As potential lightning onset triggers \citep{1999PhLA..254...79G} and as cloud nucleation seed particles \citep[e.g.,][]{1997JASTP..59.1225S}, cosmic rays can even be regarded as relevant to terrestrial climate. Fortunately, the Earth is protected from the CRF by the action of its magnetic field as well as that of the Sun. 

When a cosmic ray hits the atmosphere it produces secondary particles, including neutrons. These neutrons pass through the atmosphere, through whatever the building the neutron monitor is housed in, and penetrate the detector. In a neutron monitor, neutron sensitive proportional tubes filled with either $^{10}{\rm BF}_{3}$ or $^{3}{\rm He}$ gas, surrounded by moderator material (typically polyethylene, to reduce the energy of the secondary cosmic ray particle to about 0.025eV) and a lead target, detect near-thermal neutrons produced locally from interacting incident particles. A tube filled with either $^{10}{\rm BF}_{3}$ or $^{3}{\rm He}$ gas responds to neutrons by the exothermic reactions $^{10}{\rm B}(n,\alpha)$ -- $^{7}{\rm Li}$, or $^{3}{\rm He}(n,p)$ -- $^{3}{\rm H}$. So even though neutrons do not leave an ion trail in the proportional tube, the energetic ions resulting from absorption of a neutron by a nucleus strip electrons from neutral atoms in the tube, producing a charge, which is detected as one count. Thousands of counts per hour are detected.

\subsection{Direct Connectivity: The Solar Minimum 28-Day CRF Periodicity}\label{oscill}
\begin{figure}[!ht]
\epsscale{0.45}
\plotone{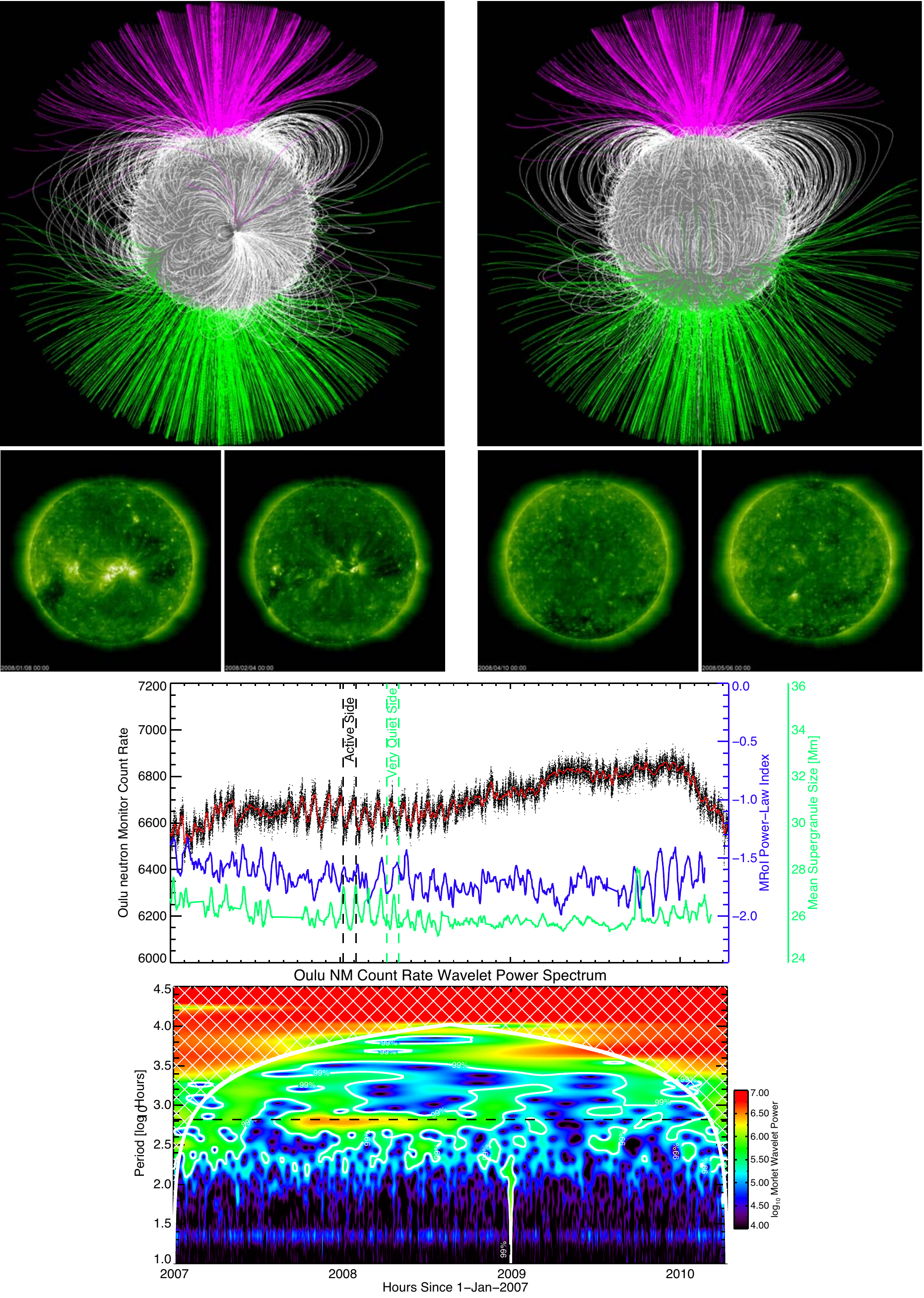}
\caption{The panel second from the bottom shows the evolution of the Oulu CRF with 6-minute sampling (black dots) and the hourly running average (red) from January 2007 to March 2010, where for comparison we show the mean scale of transition region supergranulation (blue) and the power-law index of the MRoI (green). The bottom panel shows the Wavelet transform of the CRF data along with the 95\% confidence contours (white) and a dashed horizontal line that indicates a 28 day period. The images in the upper half of the figure show the active (left central pair) and quiet (right central pair) side of the solar atmosphere and the corresponding PFSS extrapolations: green field lines are open and rooted in negative polarity field while pink field lines are rooted in positive polarity field. \label{fY}} 
\end{figure}

Often at times of solar minimum the CRF displays a strong $\sim$28 day periodicity \citep[e.g.,][]{2000SoPh..197..157B,1999ICRC....6..444K}. Figure~\pref{f2} has two examples of this behavior: the first, centered on January 1997, and the second on February 2008 (the solar cycle 22/23 and 23/24 minima) lasting around a year each. With the data shown in the panel of Fig.~\pref{fY} second from the bottom we consider the 6-minute sampling of the Oulu CRF between 2007 January~1 and 2010 March~1 (black points) with a running hourly average trace shown in red. Around 2007 September~1 the pronounced oscillation appears and has a jagged, almost sawtooth, shape---the red curve displays, on average, a steep rise and shallower descent in each period. Applying a Morlet wavelet transform \citep{1998BAMS...79...61T} to the 6-minute data (bottom panel) we see that there is considerable significant signal inside the white contours (95\% confidence intervals) at a period of 28 days (the horizontal dashed line.) The on-disk markers of small spatial scale magnetic activity---the mean supergranular scale of EIT 304\AA{} and the magnetic range of influence (MRoI) power law index \citep[measurements discussed in][]{2011ApJ...730L...3M} also vary in an oscillatory fashion during this period that are entirely consistent with an longitudinal asymmetry in the magnetic field---with a single active region in one solar hemisphere and a very quiet opposite hemisphere. Large (peak) values of the MRoI index and supergranular scale in this case are associated with the presence of equatorial coronal holes \citep{2011ApJ...730L...3M} and see the CRF drop rapidly, while the troughs of MRoI index and supergranular scale see a more gradual recovery. Further \citep[from][]{2011ApJ...730L...3M} the period of time through which the CRF begins to rise saw these length-scale measures of surface magnetism reach low levels that likely began to impact the transport of mass and energy throughout the quiescent outer solar atmosphere (including coronal holes).

The middle row of panels of Fig.~\pref{fY} illustrate the front-back asymmetric distribution of activity, with one side was dominated by a complex of the coronal holes discussed above and the last active regions of solar cycle 23 and the other side was very weak, (very) quiet sun, and as we see from Fig.~\pref{f3}, was dominated by negative polarity magnetic field. The middle row of panels in Fig.~\pref{fY} show (left) the active side of the disk on two consecutive rotations (marked by the black vertical dashed lines on the bottom panel) and (right) the quiet, negative dominant, side (marked by the green vertical lines) on the other. The very quiet hemisphere is likely the cause of the rapid increase ($\sim$100 counts) in the CRF while the additional heliospheric magnetic field from the active side leads to reduced CRF. The response time in each case is 3, or 5 days respectively, i.e., approximately the fast and slow wind travel times from the Sun to the Earth, where the difference in these times likely drives the repeating sawtooth pattern. The PFSS extrapolations on the top of Fig.~\pref{fY} illustrate the pronounced difference of the approximate coronal magnetic morphology on opposite sides of the Sun and the encroachment of open magnetic field (green traced magnetic field lines) northward in the quiet hemisphere.

The evolution of the 6-minute sampling of the Oulu CRF (black dots) and the hourly running average (red line) from January 2007 to March 2010. For comparison the blue and red traces show the mean scale of supergranulation measured in the transition region and power-law index of the MRoI describing the photospheric magnetic field element separation, respectively. These vary in an oscillatory fashion, just like the CRF. The active side of the solar disk has larger supergranules and shallower field distribution while the weak, quiet side, has smaller supergranules and steeper field distributions that are indicative of reduced spacing of magnetic elements in the photosphere. The images in the upper panels illustrate the active and quiet sides of the disk on consecutive rotations and are marked by vertical dashed lines that are black and green respectively.


\end{document}